\documentclass[sigconf]{acmart}

\usepackage{graphicx} 

\usepackage{times}
\usepackage{soul}
\usepackage{url}
\usepackage{xcolor}
\usepackage[utf8]{inputenc}
\usepackage{graphicx}
\usepackage[retainorgcmds]{IEEEtrantools} 
\usepackage{booktabs}
\usepackage{multirow}
\usepackage{bigstrut}
\usepackage{xcolor}
\usepackage{makecell}
\usepackage{subfigure}
\usepackage{algorithm}
\usepackage{algorithmicx}
\usepackage{algpseudocode}
\usepackage{appendix}

\renewcommand{\paragraph}[1]{\par\noindent\textbf{#1}.}

\AtBeginDocument{%
  }

\renewcommand\footnotetextcopyrightpermission[1]{}
\setcopyright{none}

\begin{document}

\title{Large Language Model Watermark Stealing  With \\ Mixed Integer Programming}
\author{Zhaoxi Zhang}
\authornote{Work done during the visit at Griffith University.}
\affiliation{%
  \institution{University of Technology, Sydney}
 \country{}
}

\author{Xiaomei Zhang}
\affiliation{%
  \institution{Griffith University}
  \country{}}

\author{Yanjun Zhang}
\affiliation{%
  \institution{University of Technology, Sydney}
  \country{}}
  
  \author{Leo Yu Zhang}
 \authornote{Corresponding author (leo.zhang@griffith.edu.au).}
\affiliation{%
  \institution{Griffith University}
  \country{}}
  
  \author{Chao Chen}
\affiliation{%
  \institution{Royal Melbourne Institute of Technology}
  \country{}}
  \author{Shengshan Hu}
\affiliation{%
  \institution{Huazhong University of Science and Technology}
  \country{}}
  \author{Asif Gill}
\affiliation{%
  \institution{University of Technology, Sydney}
  \country{}}
  \author{Shirui Pan}
\affiliation{%
  \institution{Griffith University}
  \country{}}



\begin{abstract}
The Large Language Model (LLM) watermark is a newly emerging technique that shows promise in addressing concerns surrounding LLM copyright, monitoring AI-generated text, and preventing its misuse. 
The LLM watermark scheme commonly includes generating secret keys to partition the vocabulary into green and red lists, applying a perturbation to the logits of tokens in the green list to increase their sampling likelihood, thus facilitating watermark detection to identify AI-generated text if the proportion of green tokens exceeds a threshold. 
However, recent research indicates that watermarking methods using numerous keys are susceptible to removal attacks, such as token editing, synonym substitution, and paraphrasing, with robustness declining as the number of keys increases. Therefore, the state-of-the-art watermark schemes that employ fewer or single keys have been demonstrated to be more robust against text editing and paraphrasing. 
In this paper, we propose a novel green list stealing attack against the state-of-the-art LLM watermark scheme and systematically examine its vulnerability to this attack. 
We formalize the attack as a mixed integer programming problem with constraints. 
We evaluate our attack under a comprehensive threat model, including an extreme scenario where the attacker has no prior knowledge, lacks access to the watermark detector API, and possesses no information about the LLM's parameter settings or watermark injection/detection scheme. 
Extensive experiments on LLMs, such as OPT and LLaMA, demonstrate that our attack can successfully steal the green list and remove the watermark across all settings.

\end{abstract}



\maketitle
\pagestyle{plain} 

\section{Introduction}

With the significant progress of Large Language Models (LLMs) in recent years~\cite{radford2019language, brown2020language,touvron2023llama, touvron2023llama2}, there are increasing risks that LLMs could be deployed for malicious purposes, such as misinformation generation~\cite{pan2023risk}, automated phishing~\cite{hazell2023large}, and academic fraud~\cite{kasneci2023chatgpt}. 
Consequently, there is a growing need to address the LLM copyright concerns, monitor AI-generated text, and prevent its misuse. 
Many existing methods involve collecting AI-generated and human-generated text and then training a classifier to distinguish them 
~\cite{Mitchell2023DetectGPTZM, Xu_Sheng_2024}. 
However, these methods tend to be biased towards the training dataset~\cite{Zhang2023WatermarksIT, pang2024attacking, jiang2023evading} and subject to adversarial attacks~\cite{krishna2023paraphrasing, sadasivan2023can, he2024can}. 

As such, LLM watermark, which enables the injection of detectable hidden patterns into the AI-generated text, emerges as a promising technique~\cite{kirchenbauer2023watermark, kirchenbauer2023reliability, zhao2023provable, liu2023semantic}. 
Typically, they inject watermarks during the text generation process, where LLMs sample the next token based on the distribution computed from the logits~\cite{kirchenbauer2023watermark, fairoze2023publicly, kirchenbauer2023reliability, ren2023robust, hou2023semstamp, liu2023semantic, zhao2023provable}. 
The watermark scheme initially generates a key
 and employs it as a seed for a pseudo-random function to randomly partition the vocabulary into a green list and a red list.  
Subsequently, a perturbation is added to the logits of tokens in the green list. 
As a result, tokens from the green list are more likely to be sampled during generation compared to those from the red list, leading to a higher frequency of green list tokens in the watermarked text. 
This can then be utilized to detect the watermark; if the proportion of green tokens exceeds a predetermined threshold, the text is considered as watermarked.  
This watermark scheme do not require modification of model parameters and can achieve high detection rates while maintaining the quality of the generated text. 

One predominant watermark approach in existing research utilizes multiple keys generated either from the token level (extracted from prefix tokens)~\cite{kirchenbauer2023watermark, kirchenbauer2023reliability, lee2023wrote} or the sentence level (obtained from sentence embeddings)~\cite{ren2023robust, hou2023semstamp, liu2023semantic}. 
However, recent studies suggest that a watermarking method employing numerous keys is vulnerable to removal attacks, such as token editing~\cite{topkara2006hiding}, synonym substitution~\cite{zellers2019defending}, and paraphrasing~\cite{sadasivan2023can, krishna2023paraphrasing}. 
Importantly, the robustness to edits deteriorates with the increasing number of keys~\cite{pang2024attacking, zhao2023provable}. Therefore, adopting fewer keys or adopting a Unigram-Watermark scheme, which only preserves one key to generate a consistent fixed green-red split, has been demonstrated to enhance the robustness to watermark removal~\cite{zhao2023provable}. 

In this paper, we demonstrate that the robustness provided by employing fewer keys or the Unigram watermark is insufficient. We present a novel watermark removal attack, wherein the attacker can steal the green list and replace the stolen green tokens with red tokens to successfully remove the watermark. 
Our stealing strategy utilizes the watermark detection rules, which provide explicit constraints that the attacker can use as guidelines. 
We model the green list stealing as a mixed integer programming problem with the objective of finding a minimal available green list for the watermark, constrained by a set of rules.

We first consider an attacker against the Unigram-Watermark scheme. We assume that they can generate text using LLMs and verifying whether the text is watermarked by querying the detector API. Additionally, they possess knowledge of the threshold employed by the watermark detector and the proportion of green tokens in the entire vocabulary. Within this setting, we first present a basic approach that sets a loose constraint on the number of green tokens by directly referencing the watermark detection threshold. We then investigate the ideal scenario with a tighter bound using an oracle method, assuming the attacker possesses precise knowledge of the number of green tokens in each sentence. Building upon this, we introduce a two-stage optimization method aimed at approximating such knowledge.

Next, we introduce an attacker without any prior knowledge: they lack access to the watermark detector API and possess no information regarding the parameter settings of the LLMs or its watermark injection/detection scheme. 
Under this setting, we present an advanced approach that can tolerate errors in the collected data. Through carefully designed constraints, this method can approach the performance of the attacker with prior knowledge. 
Furthermore, we extend this method to target watermark schemes that employ multiple keys, including both token-level and sentence-level schemes. 
To efficiently optimize additional variables and constraints introduced by the multi-key watermarks, we propose an iterative algorithm that can simultaneously steal multiple green lists  with a high true positive rate. 

Our contributions are summarized as:
\begin{itemize}
    \item We are the first work to propose a systematic watermark removal method against the state-of-the-art LLM watermark scheme. 

    
    \item We assess our attack under a comprehensive threat model including real-world attack scenarios considering varying levels of knowledge that the adversary can obtain. 
    Extensive experiments conducted on LLMs including OPT and LLaMA demonstrate the effectiveness of our attack across all settings.
    \item We release the source code and the artifact at \url{https://anonymous.4open.science/r/mip_watermark_stealing-78C9}, to facilitate future studies in this area.
\end{itemize}

\section{Background and Related Work}

\subsection{LLM Watermark}
\label{subsec:LLMwatermark}

In this section,  we formalize the problem of 
LLM watermarking. Table~\ref{tab:notation} in Appendix~\ref{app:notation} shows some important notations used in this paper.
Let $T=\{ t_j \}$ denote the vocabulary of a LLM, $t_j$ is the j-th token in vocabulary, where $|T|=m$. 
Let $S=\{S_i\}$ be the set of sentences, where $|S|=n$ and $i \in [1,n]$.   
It includes watermarked sentences (denoted as  $\hat{S}$),  and natural sentences (denoted as $\tilde{S}$),  
i.e., $S = \hat{S} \cup \tilde{S}$.


To add a watermark to an LLM, 
the model owner generates a key $k$ and use it as a seed for a pseudo-random function 
to randomly split the vocabulary into a green list $T_g$ and a red list $T_r$, where $T_g \cap T_r=\phi$. 
The proportion of the green list $T_g$ in the whole vocabulary is $\gamma$. 
A perturbation $\delta$ is then added to the logits corresponding to tokens belonging to the green list. 

One line of existing works employ multiple keys based on token level or sentence level. 
Token-level approaches generate each $k$ from the prefix tokens of length {$(q-1)$}~\cite{kirchenbauer2023watermark}, while sentence-level methods propose to generate $k$ based on the embedding of each sentence so that the watermark can process stronger robustness against adversaries such as token editing attack, synonym substitution attack and paraphrasing attack~\cite{ren2023robust, hou2023semstamp, liu2023semantic}. 
However, 
recent works demonstrate that the robustness to edits deteriorates with the increasing number of keys~\cite{zhao2023provable, pang2024attacking}.  
By employing a Unigram-Watermark scheme which uses a fixed green-red split consistently, the robustness to edits can be enhanced by twice to the existing schemes with multiple keys~\cite{zhao2023provable}. 

\subsection{Detecting Watermark}


Let 
$s_{i, j}$ 
denotes frequency of each token in a sentence, i.e., the number of token $t_j$ in $S_i$. For example, if $s_{i,j} = 5$, it means $t_j$ appears 5 times in sentence $S_i$. 
We let $l_i = |S_i|$ be the length of sentence $S_i$, and 
$s_{i,j} \in [0,l_i]$. 
We then define $C=\{c_j\}$, where $c_j \in \{0,1\}$ is the color code, i.e., $c_j=1$ and $c_j=0$ represent that  token $t_j$ belongs to the green list $T_g$ and  the red list $T_r$, respectively. 
The number of green tokens in a sentence $S_i$ 
can be computed as:
\begin{IEEEeqnarray}{c}
    G(S_i)=\sum\nolimits_{t_j \in T} s_{i,j} \cdot c_j.
    \label{eq:green_num}
\end{IEEEeqnarray}

As the proportion of green tokens in watermarked sentences is higher than the normal level,  
the commonly employed detection methods use $z$-test to evaluate the proportion of green tokens:  
\begin{equation}
    z = (G(S_i)-\gamma l_i)/\sqrt{l_i\gamma(1-\gamma)}. 
\end{equation}
If the $z$-test score exceeds the threshold, denoted by $z^*$, the sentence is considered to be watermarked. 
We then define $g_i$ as the watermark threshold of the number of green tokens for a given sentence $S_i$:
\begin{IEEEeqnarray}{c}
    g_i=z^*\sqrt{l_i\gamma(1-\gamma)}+\gamma l_i.
    \label{eq:gi}
\end{IEEEeqnarray}
Therefore,  $G(\hat{S}_i)$ should be greater than $g_i$ for a watermarked sentence $\hat{S}_i$, whereas  $G(\tilde{S}_i)$ should be less than $g_i$  for a natural sentence $\tilde{S}_i$.

\subsection{Watermark Stealing}
Existing 
watermark stealing methods are based on the token frequency to reconstruct the green list~\cite{zhao2023provable, jovanovic2024watermark, Wu2024BypassingLW}. 
As the frequency of tokens in the green list is  larger than in the red list, 
if the frequency of a token $t_j$ in the watermarked text is greater than the frequency of $t_j$ in natural text, then $t_j$ is regarded as a green token. 
However, it is difficult for frequency-based methods to differentiate low-entropy tokens, which can exhibit high frequency in both watermarked and natural text, leading to a high false positive rate and reducing the effectiveness of watermark removal. 
Also, frequency-based stealing cannot accurately identify tokens in sentences with a number of green tokens near the detection threshold. 
In addition, frequency-based methods are ineffective against multi-key watermarks, as the union of multiple green lists can encompass the entire vocabulary. 



\section{Threat Model}
We consider an attacker who aims to remove the LLM watermark by stealing the green list. 
The attack settings are based on the attacker's level of knowledge about the LLM, as detailed below.
\begin{itemize}
    \item \textbf{AS1: Watermark Detector API}. Typically, LLMs and their corresponding watermark detectors are deployed online as APIs, restricting clients to black-box access. In this setting, attackers can generate text using the LLMs and verify whether the text is watermarked by calling the detector API. They also possess the knowledge of  $\gamma$ (i.e., the proportion of the green list) and the $z$-test score threshold $z^*$. 
    \item \textbf{AS2: No-Knowledge}. In this setting,  attackers cannot access the watermark detector API. They also do not know the $z$-test score threshold or the value of $\gamma$. 
\end{itemize}

In both settings, knowledge of $\delta$ is not required.

\section{Green list stealing} 
\label{sec:steal}

\begin{figure}[t]
    \centering
    \includegraphics[width=0.48\textwidth]{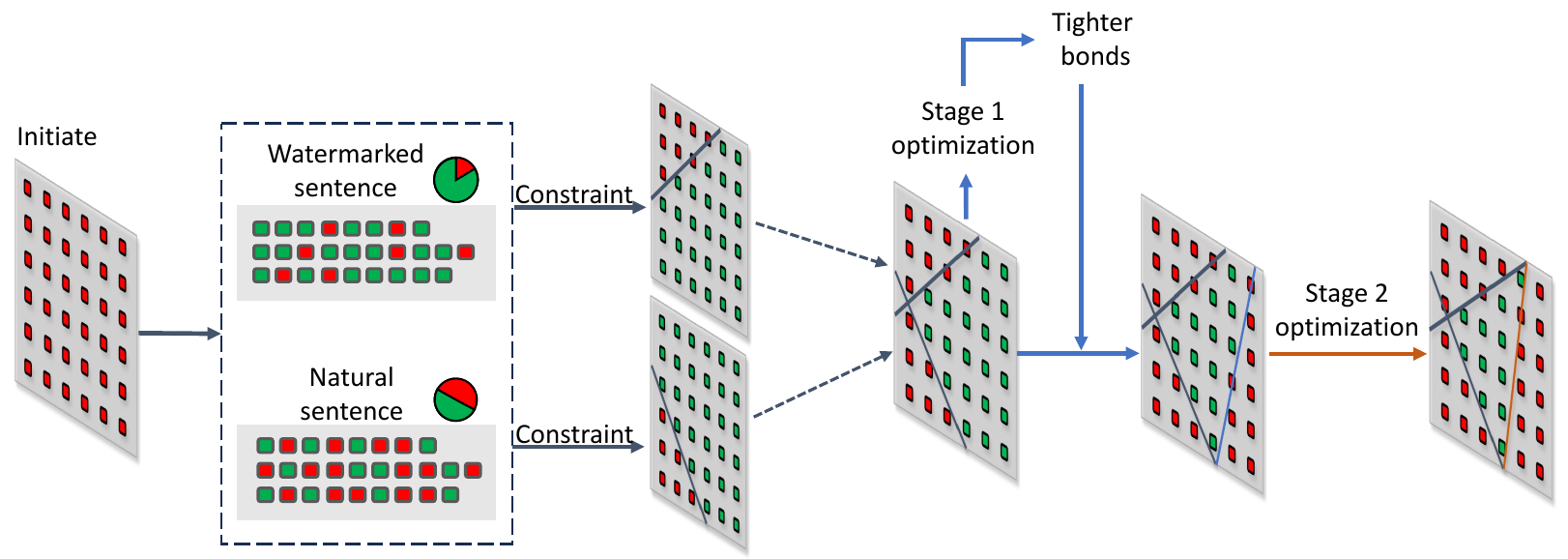}
    \caption{
    An overview of our two-stage optimization-based stealing method. The green and red squares denote the color states of tokens in the vocabulary, while the bold solid lines represent constraints. Constraints guided by watermarked sentences, natural sentences, and detection rules initially delineate the feasible region for green tokens.  Subsequently, the Stage 1 of optimization can identify tighter bounds to the feasible region.  Using these bounds in the Stage 2 of optimization, we obtain the minimal available green list.
    }
    \label{fig:steal}
\end{figure}

The insight of our stealing strategy is to utilize the watermark rules, which provide explicit constraints. 
As shown in Figure~\ref{fig:steal},  these constraints can encircle a feasible region for green tokens within the vocabulary. Stealing the green list with a high true positive rate can thus be formed as an optimization problem with the objective of finding the smallest set of green tokens that satisfy all constraints. 
Compared to frequency-based stealing, our approach naturally encompasses low entropy tokens, sentences close to the detection threshold, and green lists in multi-key watermarks as new constraints. 
Since the color of tokens can be represented as integers, we propose to steal the green lists via mixed integer programming. 

\subsection{AS1 Attacker}~\label{sec:as1}
We start with the AS1 attacker.  
First, we introduce a basic method, Vanilla-AS1, which models the green list stealing as a mixed integer programming problem constrained by the watermark threshold for the number of green tokens (Section~\ref{sec:vanilla_sce}). 
However, such constraint is loose. 
We hypothesize that a tighter bound on the number of green tokens per sentence, compared to the watermark threshold, is necessary for more accurate results. 
To this end, we explore the ideal scenario with an oracle method, Oracle-AS1, which assumes the attacker knows the exact number of green tokens in each sentence (Section~\ref{sec:oracle}). 
Building on this, we propose a two-stage optimization method, Pro-AS1, to approximate the ground truth of the number of green tokens (Section~\ref{sec:tighter_method}).

\subsubsection{\textbf{Vanilla-AS1}}
\label{sec:vanilla_sce}

In this section, we introduce a vanilla stealing method under the attack setting AS1, where the attacker has access to the watermark detector API and is aware of the $z$-test score threshold $z^*$ and the portion of green list $\gamma$. 
Attackers can only access inputs and outputs of the LLM, treating its parameters as a black box. 

Consequently, they must rely on the following characteristics of watermarked versus natural sentences: (1) The number of green tokens in a watermarked sentence exceeds the threshold. (2) The number of green tokens in a natural sentence is below the threshold. 
Based on these characteristics, we can model watermarking as a mixed integer programming problem in the following. 

\paragraph{Constraints} 
First, the number of green token in each sentence 
is constrained by the watermark threshold $g_i$ (c.f., Eq.~(\ref{eq:gi})): 
\begin{IEEEeqnarray}{c}
\begin{aligned}
    \mathrm{G}(S_i) \geq g_i , \forall S_i \in \hat{S},\\
    \mathrm{G}(S_i) \leq g_i , \forall S_i \in \tilde{S}.
    \label{eq:boundary_native}
\end{aligned}
\end{IEEEeqnarray}
In addition, the number of green tokens should be less than $\gamma|T|$,  and this can be formulated as follows:
\begin{IEEEeqnarray}{c}
    \sum_{t_j \in T} c_{j} \leq \gamma |T|.
    \label{eq:sum_token_num_native}
\end{IEEEeqnarray}

\paragraph{Objective Function} 
To reduce false positives in the stolen green list, the objective of the attacker is to find a minimal viable green list while satisfying the constraints in Eq.~(\ref{eq:boundary_native}) and Eq.~(\ref{eq:sum_token_num_native}): 
\begin{IEEEeqnarray}{l}
\begin{aligned}
    \mathrm{minimize} &~ \sum_{t_j \in T} c_{j} \cdot w_j,\\
    \mathrm{subject \ to \ } &\mathrm{Eq.}~ 
    (\ref{eq:boundary_native}),(\ref{eq:sum_token_num_native}),
    \label{eq:vanilia_method}
\end{aligned}
\end{IEEEeqnarray}
where $w_j$ is the weight of each token in the vocabulary. It represents the ratio of a token's frequency in natural sentences to its frequency in watermarked sentences, calculated as:  
\begin{IEEEeqnarray}{c}
    w_j = \frac{\mathrm{Frequency}(t_j \mathrm{\ in \ \tilde{S}})}{\mathrm{Frequency}(t_j \mathrm{\ in \ \hat{S}})}.
    \label{eq:frequency_weight}
\end{IEEEeqnarray}
Intuitively, adding the weights will steer the optimization to remove tokens with higher $w_j$ (i.e., those appearing more frequently in natural sentences) while retaining tokens with lower $w_j$ (i.e., those appearing more frequently in watermarked sentences). 



\subsubsection{\textbf{Oracle-AS1}}
\label{sec:oracle}
In Vanilla-AS1, the constraints of Eq.~(\ref{eq:boundary_native})  only rely on the  watermark threshold $g_i$. These constraints are relatively loose, whereas imposing a tighter bound on the number of green tokens in a sentence can lead to more precise constraints for the integer programming problem to yield better convergence performance. 
In this section, we investigate the best case scenario in which the attacker is able to obtain  the exact number of green tokens in each sentence.


\paragraph{Constraints} We define $\hat{g}^o_i$ and $\tilde{g}^o_i$ as the ground-truth of the number of green tokens in the sentences from $\hat{S}$ and $\tilde{S}$, respectively. 
Then, the inequalities constraints of Eq.~(\ref{eq:boundary_native}) can be written as: 
\begin{IEEEeqnarray}{c}
\begin{aligned}
    \mathrm{G}(S_i) \geq \hat{g}^o_i, \forall S_i \in \hat{S},\\
    \mathrm{G}(S_i) \leq \tilde{g}^o_i, \forall S_i \in \tilde{S}.
    \label{eq:wm_nl_boundary_o}
\end{aligned}
\end{IEEEeqnarray}
\paragraph{Objective Function} Correspondingly, the objective function can be transformed into:
\begin{IEEEeqnarray}{lcl}
\begin{aligned}
    \mathrm{minimize} &~ \sum_{t_j \in T} c_{j} \cdot w_j,\\
    \mathrm{subject \ to} &~ \mathrm{Eq.}~
    (\ref{eq:wm_nl_boundary_o}), (\ref{eq:sum_token_num_native}).
    \label{eq:oracle_method}
\end{aligned}
\end{IEEEeqnarray}
Compared to Eq.~(\ref{eq:boundary_native}) in Vanilla-AS1, Eq.~(\ref{eq:wm_nl_boundary_o}) imposes more precise constraints, which is expected to result in a more powerful attack with a higher true positive rate. 

\subsubsection{\textbf{Pro-AS1}}
\label{sec:tighter_method}
In this section, we eliminate the assumption of knowing the exact number of green tokens in Oracle-AS1. 
Instead, we propose a two-stage optimization method that allows the attacker to approximate the exact number of green tokens. 

\paragraph{Stage 1} 
In this stage, the attacker aims to estimate the number of green tokens. 
We establish bounds $\hat{b}_i$ and $\tilde{b}_i$ for watermarked and natural sentences, respectively, to substitute $\hat{g}^o_i$ and $\tilde{g}^o_i$ in Eq.~(\ref{eq:wm_nl_boundary_o}). 
\paragraph{Constraints} The constraints of the Stage 1 can be described as:
\begin{IEEEeqnarray}{rcl}
    \mathrm{G}(S_i) \geq \hat{b}_i &, \ & \forall S_i \in \hat{S},
    \label{eq:wm_boundary_plus1}\\
    \hat{b}_i       \geq g_i &, \ & \forall S_i \in \hat{S},
    \label{eq:wm_boundary_plus2}\\
    \mathrm{G}(S_i) \leq \tilde{b}_i &, \ & \forall S_i \in \tilde{S},
    \label{eq:nl_boundary_plus1}\\
    \tilde{b}_i     \leq g_i &, \ & \forall S_i \in \tilde{S}.
    \label{eq:nl_boundary_plus2}
\end{IEEEeqnarray}

\paragraph{Objective Function} 
To approximate the  the exact number of green tokens, 
we maximize $\hat{b}_i$ for each watermarked sentence to increase the number of green tokens therein as much as possible, while ensuring that the number of green tokens in natural sentences remains close to the average level. 
The objective function is presented as follows:
\begin{IEEEeqnarray}{lcl}
\begin{aligned}
    \mathrm{maximize} &~ \sum_{S_i \in \hat{S}} \hat{b}_i - \mathrm{abs}(\sum_{S_i \in \tilde{S}} \tilde{b}_i-\gamma \cdot \sum_{S_i \in \tilde{S}}l_i),\\
    \mathrm{subject \ to} &~ \mathrm{Eq.}~
    (\ref{eq:wm_boundary_plus1}),(\ref{eq:wm_boundary_plus2}),(\ref{eq:nl_boundary_plus1}),(\ref{eq:nl_boundary_plus2}),(\ref{eq:sum_token_num_native}),
    \label{eq:find_boundary_abs_tmp}
\end{aligned}
\end{IEEEeqnarray}
where $l_i$ is the length of sentence $S_i$. 
Due to the non-linearity introduced by the absolute value in Eq.~(\ref{eq:find_boundary_abs_tmp}), this objective function cannot be directly optimized using mixed integer programming. Therefore, we introduce an equivalent variable, $\tilde{b}^{(abs)}$, to replace the absolute value:

\begin{IEEEeqnarray}{rcl}
\begin{aligned}
    \tilde{b}^{(abs)} &\geq& \sum_{S_i \in \tilde{S}} \tilde{b}_i-\gamma \cdot \sum_{S_i \in \tilde{S}}l_i,\\
    \tilde{b}^{(abs)} &\geq& -\sum_{S_i \in \tilde{S}} \tilde{b}_i+\gamma \cdot \sum_{S_i \in \tilde{S}}l_i.
    \label{eq:abs_bi}
\end{aligned}
\end{IEEEeqnarray}
As such, Eq.~(\ref{eq:find_boundary_abs_tmp}) can be written as: 
\begin{IEEEeqnarray}{lcl}
\begin{aligned}
    \mathrm{maximize} &~ \sum_{S_i \in \hat{S}} \hat{b}_i - \tilde{b}^{(abs)},\\
    \mathrm{subject \ to} &~ \mathrm{Eq.}~
    (\ref{eq:wm_boundary_plus1}),(\ref{eq:wm_boundary_plus2}),(\ref{eq:nl_boundary_plus1}),(\ref{eq:nl_boundary_plus2}), (\ref{eq:abs_bi}), (\ref{eq:sum_token_num_native}).
    \label{eq:find_boundary_abs}
\end{aligned}
\end{IEEEeqnarray}

\begin{figure}[t]
    \centering
    \includegraphics[height=3.3cm]{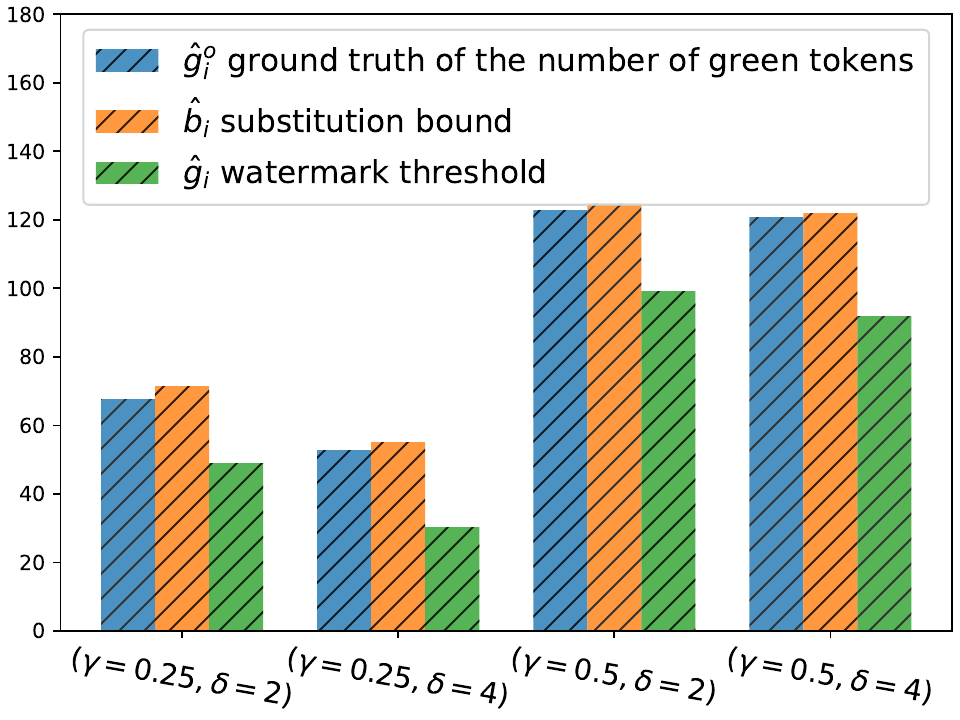}
    \caption{
    The ground truth of the number of green tokens $\hat{g}^o_i$ in a watermarked sentence is significantly greater than the watermark threshold $g_i$ used in Vanilla-AS1, resulting in loose constraints. 
    The substitution bound $\hat{b}_i$ found by Pro-AS1 can approximate $\hat{g}^o_i$, providing tighter constraints. 
    The victim model is OPT-1.3B, and similar phenomena are observed in other sentences. 
    }
    \label{fig:g_i_b_i}
\end{figure}

Figure~\ref{fig:g_i_b_i} shows the performance of Pro-AS1 in approximating the ground truth of the number of green tokens. While the difference between $g_i$ used in Vanilla-AS1 and the ground truth $\hat{g}^o_i$ is larger than 20, the difference between $\hat{b}_i$ found by Pro-AS1 and $\hat{g}^o_i$ is less than 5. This minor discrepancy between $\hat{b}_i$ and $\hat{g}^o_i$  arises from the impracticality of achieving the global optimum within a finite time on a large dataset. 
 
\paragraph{Stage 2} 
In this stage, the attacker aims to obtain the green tokens. 
We let $\hat{b}_{sum} = \sum_{S_i \in \hat{S}} \hat{b}_i$ and $\tilde{b}_{sum} = \sum_{S_i \in \tilde{S}} \tilde{b}_i$ derived by optimization in Eq.~(\ref{eq:find_boundary_abs}). 
To tolerate the minor discrepancy between $b_i$ and $\hat{g}^o_i$, we introduce hyperparameters $\hat{\beta}$ and $\tilde{\beta}$ to rescale the bounds. 
\paragraph{Constraints}
As such, the constraints of the bound of the number of the green tokens in watermarked sentences and natural sentences can be formed as 
\begin{IEEEeqnarray}{rcl}
\begin{aligned}
    \sum_{S_i \in \hat{S}} \hat{b}_i &\geq \hat{\beta} \cdot \hat{b}_{sum},\\
    \sum_{S_i \in \tilde{S}} \tilde{b}_i &\leq \tilde{\beta} \cdot \tilde{b}_{sum}.
    \label{eq:sum_boundary_plus}
\end{aligned}
\end{IEEEeqnarray}
\paragraph{Objective Function} 
Incorporating the constraints in Eq.~(\ref{eq:sum_boundary_plus}), the objective function in the Stage 2 is represented as:
\begin{IEEEeqnarray}{lcl}
\begin{aligned}
    \mathrm{minimize} &~ \sum_{t_j \in T} c_{j} \cdot w_j,\\
    \mathrm{subject \ to} &~ \mathrm{Eq.}~
    (\ref{eq:wm_boundary_plus1}),(\ref{eq:wm_boundary_plus2}),(\ref{eq:nl_boundary_plus1}),(\ref{eq:nl_boundary_plus2}), (\ref{eq:sum_boundary_plus}), (\ref{eq:sum_token_num_native}).
    \label{eq:pro_method}
\end{aligned}
\end{IEEEeqnarray}


\subsection{AS2 Attacker}

In this section, we discuss how to extend Pro-AS1 from Section~\ref{sec:tighter_method} to the AS2 attack setting. 
Similar to Pro-AS1, the AS2 attacker also uses a two-stage optimization process: finding the number of green tokens in the Stage 1 and obtaining the green list in the Stage 2.



\paragraph{Stage 1} 
The AS2 attacker does not have access to the watermark detector API, thus they cannot verify whether a sentence is watermarked or not. 
The attacker can only assume that all sentences generated by the LLM are watermarked, while those obtained from the wild are unwatermarked (natural). 
However, because watermarking relies on sampling, there are instances where the LLM fails to apply the watermark to its output. Additionally, natural text can be erroneously identified as watermarked by the watermark detector. 
Due to a lack of verification by the watermark detector API, two types of erroneous samples emerge: (1) the LLM output lacks the watermark, and (2) natural text is incorrectly labeled as watermarked.

Handling the erroneous samples is necessary; otherwise, they will make the solution of the mixed integer programming infeasible.  
To address this, we introduce binary variables $\lambda_i \in \{0,1\}$ to determine whether sentence $S_i$ should be included into the optimization. 
Specifically, $\lambda_{i}=1$  indicates that  sentence $S_i$ is not erroneous and  should be considered, 
while $\lambda_{i}=0$  indicates that  sentence $S_i$ is erroneous and  should be disregarded during optimization. 
\paragraph{Constraints}
After incorporating $\lambda_i$ into Eq.~(\ref{eq:wm_boundary_plus1}) and (\ref{eq:nl_boundary_plus1}), new constraints are defined as follows:
\begin{IEEEeqnarray}{c}
    \mathrm{G}(S_i) \geq (\hat{b}_i + (\lambda_{i}-1)\cdot l_i ), \forall S_i \in \hat{S},
    \label{eq:wm_boundary_noapi}\\
    \mathrm{G}(S_i) \leq (\tilde{b}_i + (1-\lambda_{i})\cdot l_i), \forall S_i \in \tilde{S}. 
    \label{eq:nl_boundary_noapi}
\end{IEEEeqnarray}
When $\lambda_{i}=1$, Eq.~(\ref{eq:wm_boundary_noapi}) and (\ref{eq:nl_boundary_noapi}) are equivalent to Eq.~(\ref{eq:wm_boundary_plus1}) and (\ref{eq:nl_boundary_plus1}). 
When $\lambda_{i}=0$, the right side of Eq.~(\ref{eq:wm_boundary_noapi}) becomes $\hat{b}_i-l_i$, which turns negative as $\hat{b}_i$ should be smaller than $l_i$. 
Since $\mathrm{G}(S_i) \geq 0$, Eq.~(\ref{eq:wm_boundary_noapi}) always holds. 
Also, when $\lambda_{i}=0$, Eq.~(\ref{eq:nl_boundary_noapi})  always holds
as $\mathrm{G}(S_i)$ is always less than $l_i$.
Therefore,  $S_i$ will be excluded from the optimization on the constraints of Eq.~(\ref{eq:wm_boundary_noapi}) and (\ref{eq:nl_boundary_noapi}). 

Similarly, we also need to introduce constraints on $\hat{b}_i$ and $\tilde{b}_i$ to handle erroneous samples. 
When  $S_i$ is  erroneous, $\hat{b}_i$ and $\tilde{b}_i$ should be set to $0$. 
Otherwise, their values can disrupt the optimization process.
For $i \in [1, n]$, the constraints on $\hat{b}_i$ and $\tilde{b}_i$ are given as: 
\begin{IEEEeqnarray}{c}
\begin{aligned}
    \lambda_{i} \cdot l_i &\geq \hat{b}_i, \forall S_i \in \hat{S},\\
    \lambda_{i} \cdot l_i &\geq \tilde{b}_i, \forall S_i \in \tilde{S}.
    \label{eq:sentence_num_noapi}
\end{aligned}
\end{IEEEeqnarray}
As $ \lambda_{i} \in \{0,1\}$ and $\hat{b}_i$, $\tilde{b}_i \geq 0$,  when $\lambda_i=0$, $\hat{b}_i$ and $\tilde{b}_i$ are set to 0 by Eq.~(\ref{eq:sentence_num_noapi}).

Also, we need to bound the number of erroneous samples as:  
\begin{IEEEeqnarray}{c}
\begin{aligned}
    p_l |\hat{S}| \leq \sum_{S_i \in \hat{S}} \lambda_{i} \leq p_u |\hat{S}|, \\
    p_l |\tilde{S}| \leq \sum_{S_i \in \tilde{S}} \lambda_{i} \leq p_u |\tilde{S}|,
    \label{eq:wheter_sentence_noapi}
\end{aligned}
\end{IEEEeqnarray}
where $p_u$ and $p_l$ are hyperparameters that represent the upper and lower bounds of erroneous samples, respectively.

We let $r_c$ as the proportion of the erroneous samples in a dataset. 
Typically, the proportion of erroneous samples is under $10\%$. However, if the LLM implements countermeasures, this proportion can significantly increase; for example, LLMs may refresh their green list occasionally, or the dataset collected by the attacker is poisoned.
When $r_c$ is very large, it will be hard for Eq.~(\ref{eq:wm_boundary_noapi}) and (\ref{eq:nl_boundary_noapi}) to find suitable $\lambda$ for each sentence. 
In this case, even within the watermarked dataset, the number of red tokens will exceed the number of green tokens. To further identify watermarked sentences, we need to rely on two characteristics of the watermark: 
(1) To avoid false positives during detection, the size of green lists is usually fewer than the red lists; 
(2) In watermarked sentences, the proportion of green tokens is usually higher than in natural sentences.
These two characteristics can be formalized as constraints as follows: 
\begin{IEEEeqnarray}{l}
\begin{aligned}
    \sum_{t_j \in T} c_{j} & \leq  \eta |T|,\\
    \min_{S_i \in \hat{S}}(G(S_i)/l_i) &-\max_{S_i \in \tilde{S}}(G(S_i)/l_i) \geq \epsilon,
    \label{eq:erroneous}
\end{aligned}
\end{IEEEeqnarray}
where $\eta$ is expected size of stolen green list, $\epsilon$ is a threshold to distinguish watermarked and natural sentence. 
It is worth noting that if the dataset size is relatively small, $\eta$ can be set as a very low value. 


\paragraph{Objective Function} 
The AS2 attacker lack knowledge of $\gamma$ for watermarking and the $z$-test score threshold $z^*$. 
Therefore, instead of maximizing $-\mathrm{abs}(\sum_{S_i \in \tilde{S}} \tilde{b}_i-\gamma \cdot \sum_{S_i \in \tilde{S}}l_i)$ in Eq.~(\ref{eq:find_boundary_abs}), our objective is changed to maximize the number of green tokens in watermarked sentences while minimizing it as much as possible in natural sentences. 
In addition, without knowing $\gamma$ and $z^*$, Eq.~(\ref{eq:wm_boundary_plus2}), (\ref{eq:nl_boundary_plus2}) also need to be discarded in the optimization. 
The objective function thus becomes
\begin{IEEEeqnarray}{lcl}
\begin{aligned}
    \mathrm{maximize} &~ \sum_{S_i \in \hat{S}} \hat{b}_i - \sum_{S_i \in \tilde{S}} \tilde{b}_i, \\
    \mathrm{subject \ to} &~ \mathrm{Eq.}~
    (\ref{eq:wm_boundary_noapi}),(\ref{eq:nl_boundary_noapi}), (\ref{eq:wheter_sentence_noapi}), (\ref{eq:sentence_num_noapi}).
    \label{eq:find_boundary_noapi}
\end{aligned}
\end{IEEEeqnarray}
\paragraph{Stage 2} 
After finding suitable $\hat{b}_i$ and $\tilde{b}_i$ by Eq.~(\ref{eq:find_boundary_noapi}), the attacker starts the Stage 2 optimization to obtain the green list.
\paragraph{Constraints} Similar to Pro-AS1, the AS2 attacker also adopts the constraints of Eq.~(\ref{eq:sum_boundary_plus}). In addition, it is constrained by Eq.~(\ref{eq:wm_boundary_noapi}), 
(\ref{eq:nl_boundary_noapi}), (\ref{eq:wheter_sentence_noapi}), (\ref{eq:sentence_num_noapi}) to handle the erroneous samples. 
\paragraph{Objective Function} 
The Stage 2 optimization is as follows: 
\begin{IEEEeqnarray}{lcl}
\begin{aligned}
    \mathrm{minimize} &~ \sum_{t_j \in T} c_{j} \cdot w_j,\\
    \mathrm{subject \ to} &~ \mathrm{Eq.}~
    (\ref{eq:wm_boundary_noapi}),(\ref{eq:nl_boundary_noapi}),(\ref{eq:wheter_sentence_noapi}), (\ref{eq:sentence_num_noapi}),
    (\ref{eq:sum_boundary_plus}).
    \label{eq:main_method}
\end{aligned}
\end{IEEEeqnarray}
Finally, the green list $T^s_g$ and red list $T^s_r$ can be stolen by traversing the values of $c_{j}$. 

\section{Multi-key Stealing}
\label{sec:multi_key}
In this section, we generalize our method from Section~\ref{sec:steal} (i.e., stealing the green list from a unigram watermark scheme that uses a fixed green-red split) to a multi-key watermark scheme (where different keys correspond to different green-red splits). The attacker is under the AS2 setting. 

As mentioned in Section~\ref{subsec:LLMwatermark}, the watermark robustness to edits deteriorates with the increasing number of keys~\cite{zhao2023provable, pang2024attacking}. As such, to make watermark immune to paraphrase attacks, the number of keys is less than {$5$ in real applications to our best knowledge~\cite{ren2023robust, hou2023semstamp}.}
Without loss of generality, we denote $K=\{k\}$ the set of keys, and in total there are $p=|K|$ keys. For each key $k \in K$, there is an associated green-red split, $T_{g_k}$ and $T_{r_k}$, and a key-dependent color code set $C_k=\{c^k_j\}$. In contrast to the unigram-watermark green list stealing, now the attacker's goal is to recover $\bigcup_{k \in K} T_{g_k}$ (or equivalently $\{c^k_j\}$). The same to the AS2 attacker in the unigram case, the attacker can only assume that all corpus generated by the LLM are watermarked, while those obtained from the wild are unwatermarked (natural). However, it is challenging for the attacker to confirm which corpus is associated with which key. 

It is worth mentioning that this multi-key stealing formulation encompasses both the token-level multi-key watermark~\cite{kirchenbauer2023watermark} and the sentence-level multi-key watermark~\cite{ren2023robust}; the only difference is whether the corpus is collected vertically or horizontally. For ease of presentation, we focus on the sentence-level case. 
For each sentence $S_i$, we use ${\rho^k_i} \in \{0,1\}$ to denote which key is suitable for this sentence. 
\paragraph{Constraints}
Considering there is only one key for each sentence, $\rho_i^k$ should be constrained by the following equation:
\begin{IEEEeqnarray}{c}
    \sum_{k\in K} \rho_i^k=1, \forall S_i \in 
    \hat{S}.
    \label{eq:sum_rho}
\end{IEEEeqnarray}
Similar to Eq.~(\ref{eq:wm_boundary_noapi}), (\ref{eq:nl_boundary_noapi}), the number of green token for sentence $S_i$ under the key $k$, $\mathrm{G}(S_i, k)$, should obey the following:
\begin{IEEEeqnarray}{lcl}
    \mathrm{G}(S_i, k) &\geq& (\hat{b}^k_i + (\rho_i^k-1+\lambda_{i}-1)\cdot l_i ), \forall S_i \in \hat{S}, k \in K,
    \label{eq:wm_boundary_multi_tmp}\\
    \mathrm{G}(S_i,k) &\leq& (\tilde{b}^k_i + (1-\lambda_{i})\cdot l_i), \forall S_i \in \tilde{S}, k \in K, 
    \label{eq:nl_boundary_multi_tmp}
\end{IEEEeqnarray}
where $\hat{b}^k_i$ and $\tilde{b}^k_i$ are the bounds for sentence $S_i$ under key $k$. And similar to Eq.~(\ref{eq:sentence_num_noapi}), $\hat{b}^k_i$ and $\tilde{b}^k_i$ should be constrained as follows:
\begin{IEEEeqnarray}{c}
\begin{aligned}
    \rho_i^k \cdot l_i \geq \hat{b}^k_i, \forall S_i \in \hat{S},\\
     l_i \geq \tilde{b}^k_i, \forall S_i \in \tilde{S}.
    \label{eq:rho_bonds}
\end{aligned}
\end{IEEEeqnarray}
Following the definition in Section~\ref{sec:tighter_method}, we let $\hat{b}_i$ and $\tilde{b}_i$ as bounds for watermarked and natural sentences under all keys, i.e.,  $\hat{b}_i=\max\nolimits_k(\hat{b}_i^k)$ and $\tilde{b}_i=\max\nolimits_k(\tilde{b}_i^k)$. 
Incorporating theses bounds into Eq.~(\ref{eq:wm_boundary_multi_tmp}) and (\ref{eq:nl_boundary_multi_tmp}), we get the following new constraints: 
\begin{IEEEeqnarray}{rcl}
    \mathrm{G}(S_i, k) &\geq& (\hat{b}_i + (\rho_i^k-1+\lambda_{i}-1)\cdot l_i ), \forall S_i \in \hat{S}, k \in K, 
    \label{eq:wm_boundary_multi}\\
    \mathrm{G}(S_i,k) &\leq& (\tilde{b}_i + (1-\lambda_{i})\cdot l_i), \forall S_i \in {\tilde{S}}, k \in K.
    \label{eq:nl_boundary_multi}
\end{IEEEeqnarray}
 
\paragraph{Objective Function} The two-stage optimization for multi-key watermark stealing is:
\begin{IEEEeqnarray}{lcl}
\begin{aligned}
    \mathrm{maximize} &~ \sum_{S_i \in \hat{S}} \hat{b}_i - \sum_{S_i \in \tilde{S}} \tilde{b}_i, \\
    \mathrm{subject \ to} &~ \mathrm{Eq.}~
    (\ref{eq:wm_boundary_multi}),(\ref{eq:nl_boundary_multi}), 
    (\ref{eq:rho_bonds}), 
    (\ref{eq:sum_rho}), (\ref{eq:wheter_sentence_noapi}), (\ref{eq:sentence_num_noapi}); 
    \label{eq:find_boundary_multi}
\end{aligned}
\end{IEEEeqnarray}
\begin{IEEEeqnarray}{lcl}
\begin{aligned}
    \mathrm{minimize} &~ \sum_{k \in K}\sum_{t_j \in T} c^k_{j},\\
    \mathrm{subject \ to} &~ \mathrm{Eq.}~
    (\ref{eq:wm_boundary_multi}),(\ref{eq:nl_boundary_multi}), 
    (\ref{eq:rho_bonds}), 
    (\ref{eq:sum_rho}),  
    (\ref{eq:wheter_sentence_noapi}), (\ref{eq:sentence_num_noapi}), (\ref{eq:sum_boundary_plus}).
    \label{eq:multi_method}
\end{aligned}
\end{IEEEeqnarray}
In the multi-key watermark setting, there exist multiple green lists whose union constitutes the entire vocabulary. So, the frequency of each token cannot accurately reflect its importance. Therefore, token weights were not considered during optimization in Eq.~(\ref{eq:multi_method}).

It is worth noting that the core idea of Eq.~(\ref{eq:find_boundary_multi}) is the same with Eq.~(\ref{eq:find_boundary_abs_tmp}) of Pro-AS1. 
However, considering $ \hat{b}_i=\max\nolimits_k(\hat{b}_i^k)$, maximizing $\sum\nolimits_{S_i \in \hat{S}} \hat{b}_i$ is a Max-Max problem, and it involves too many bool variables $\rho_i^k$. 
This makes it very hard to converge in mixed integer programming. 
To handle this problem, we propose an iterative method, as shown in Algorithm~\ref{alg:multi_key}. 
In the beginning, we randomly initialize $\{\rho_i^k\}$ to assign each sentence a key. 
In each iteration of the algorithm, we first fix $\{\rho_i^k\}$. Then, we use a two-stage optimization to adjust the remaining variables. Finally, after optimization, we reassign the most suitable key to each sentence based on the results of $C_k$:  
\begin{IEEEeqnarray}{c}
    \rho_i^k=
    \begin{cases}
        1, \textrm{if} \ k=\mathrm{arg}\max\limits_{k}(\mathrm{G}(S_i, k) );\\
        0, \textrm{else}.
    \end{cases}
\label{eq:rho}
\end{IEEEeqnarray}
To prevent the optimization from finding green lists with small size and getting stuck in local optima during the early stages of the iterative algorithm, we add the following constraint into Eq.~(\ref{eq:find_boundary_multi}) and (\ref{eq:multi_method}) to limit the size of the green list:
\begin{IEEEeqnarray}{c}
     \sum_{t_j \in T} c^k_{j} \geq \mu, \forall k \in K, 
    \label{eq:mu}
\end{IEEEeqnarray}
where $\mu$ is a hyperparameter to limit the size of green lists.

\begin{algorithm}
    \caption{
    Stealing green lists in multi-key watermark scheme
    }
    \label{alg:multi_key}
    \begin{algorithmic}[1]
        \Require $C_k=\{c_j^k\}, c_j^k\in \{0,1\}$, $\rho_i^k \in \{0,1\}$, $k \in K$
        \State Random initialize $\{\rho_i^k\}$. 
        \For {until optimization converge}
            \State Optimize Eq.~(\ref{eq:find_boundary_multi}) while fixing $\rho_i^k$;
            \State Optimize Eq.~(\ref{eq:multi_method}) while fixing $\rho_i^k$;
            \State Resign $\rho_i^k$ according to Eq.~(\ref{eq:rho}); 
        \EndFor
    \end{algorithmic}
\end{algorithm}

\begin{table*}[]
  \centering
  \caption{
  AS1 attacker performance of green list stealing against LLaMA-2-7B. $N_g$ and $N_t$ represent the number of green tokens and the number of true green tokens, respectively. Precision = $N_t$/$N_g$.
  }
  \resizebox{0.75\textwidth}{!}{
    \begin{tabular}{cr|rrr|rrr|rrr|rrr}
    \toprule
    \multirow{2}[2]{*}{\makecell{Watermark\\Setting}} & \multicolumn{1}{c|}{\multirow{2}[2]{*}{\makecell{Dataset\\Size}}} & \multicolumn{3}{c|}{Vanilla} & \multicolumn{3}{c|}{Oracle} & \multicolumn{3}{c|}{Pro} & \multicolumn{3}{c}{Freq.} \\
          &       & \multicolumn{1}{c}{$N_g$} & \multicolumn{1}{c}{$N_t$} & \multicolumn{1}{c|}{Precision$(\uparrow)$} & \multicolumn{1}{c}{$N_g$} & \multicolumn{1}{c}{$N_t$} & \multicolumn{1}{c|}{Precision$(\uparrow)$} & \multicolumn{1}{c}{$N_g$} & \multicolumn{1}{c}{$N_t$} & \multicolumn{1}{c|}{Precision$(\uparrow)$} & \multicolumn{1}{c}{$N_g$} & \multicolumn{1}{c}{$N_t$} & \multicolumn{1}{c}{Precision$(\uparrow)$} \\
    \midrule
    \multirow{4}[2]{*}{\makecell{$\gamma=0.25$\\$\delta=2$}} & 4000  & 662   & 537   & 81.12$\%$ & 3326  & 2798  & 84.13$\%$ & 1064  & 885   & 83.18$\%$ & 5154  & 2547  & 49.42$\%$ \\
          & 10000 & 1087  & 918   & 84.45$\%$ & 4081  & 3942  & 96.59$\%$ & 1431  & 1224  & 85.53$\%$ & 5519  & 2970  & 53.81$\%$ \\
          & 20000 & 1604  & 1351  & 84.23$\%$ & 4473  & 4408  & 98.55$\%$ & 1396  & 1256  & 89.97$\%$ & 5494  & 3181  & 57.90$\%$ \\
          & 40000 & 2749  & 2003  & 72.86$\%$ & 4778  & 4740  & 99.20$\%$ & 2146  & 1912  & 89.10$\%$ & 5425  & 3335  & 61.47$\%$ \\
    \midrule
    \multirow{4}[2]{*}{\makecell{$\gamma=0.25$\\$\delta=4$}} & 4000  & 554   & 513   & 92.60$\%$ & 3491  & 3282  & 94.01$\%$ & 732   & 678   & 92.62$\%$ & 4350  & 2867  & 65.91$\%$ \\
          & 10000 & 763   & 706   & 92.53$\%$ & 4138  & 4069  & 98.33$\%$ & 780   & 731   & 93.72$\%$ & 4704  & 3259  & 69.28$\%$ \\
          & 20000 & 934   & 869   & 93.04$\%$ & 4510  & 4473  & 99.18$\%$ & 867   & 803   & 92.62$\%$ & 4895  & 3498  & 71.46$\%$ \\
          & 40000 & 1111  & 1027  & 92.44$\%$ & 4860  & 4834  & 99.47$\%$ & 933   & 861   & 92.28$\%$ & 5020  & 3737  & 74.44$\%$ \\
    \midrule
    \multirow{4}[2]{*}{\makecell{$\gamma=0.5$\\$\delta=2$}} & 4000  & 1747  & 1527  & 87.41$\%$ & 6199  & 5554  & 89.60$\%$ & 2136  & 1884  & 88.20$\%$ & 6417  & 4784  & 74.55$\%$ \\
          & 10000 & 2426  & 2123  & 87.51$\%$ & 7943  & 7792  & 98.10$\%$ & 2253  & 2035  & 90.32$\%$ & 7233  & 5643  & 78.02$\%$ \\
          & 20000 & 3665  & 3187  & 86.96$\%$ & 8753  & 8677  & 99.13$\%$ & 2633  & 2394  & 90.92$\%$ & 7661  & 6152  & 80.30$\%$ \\
          & 40000 & 4625  & 4065  & 87.89$\%$ & 9353  & 9317  & 99.62$\%$ & 3245  & 2976  & 91.71$\%$ & 7811  & 6460  & 82.70$\%$ \\
    \midrule
    \multirow{4}[2]{*}{\makecell{$\gamma=0.5$\\$\delta=4$}} & 4000  & 1646  & 1521  & 92.41$\%$ & 6618  & 6261  & 94.61$\%$ & 2204  & 2047  & 92.88$\%$ & 6240  & 5211  & 83.51$\%$ \\
          & 10000 & 2196  & 2031  & 92.49$\%$ & 8139  & 8031  & 98.67$\%$ & 3308  & 3078  & 93.05$\%$ & 7351  & 6242  & 84.91$\%$ \\
          & 20000 & 2701  & 2498  & 92.48$\%$ & 8868  & 8827  & 99.54$\%$ & 3398  & 3174  & 93.41$\%$ & 7855  & 6792  & 86.47$\%$ \\
          & 40000 & 3589  & 3308  & 92.17$\%$ & 9454  & 9430  & 99.75$\%$ & 3533  & 3336  & 94.42$\%$ & 8173  & 7205  & 88.16$\%$ \\
    \bottomrule
    \end{tabular}%
  \label{tab:steal_as1_llama}%
  }
\end{table*}%

\section{Green Token Removing}

We now employ watermark removal based on the stolen $T^s_g$ and $T^s_r$. 
We adopt two removal strategies in this paper, i.e., greedy search-based method and gumbel softmax-based method. 

\subsection{Greedy Search}
We first consider replacing tokens from the stolen green list with the most similar tokens not in the stolen green list, i.e., for $t_j \in T^s_g$, 
\begin{IEEEeqnarray}{c}
    T^c_j=\mathrm{F_s}(t_j) \cap T^s_r,
\end{IEEEeqnarray}
where $F_s$ is the function to generate synonyms of input tokens, $T^s_r$ is the stolen red list, and $T^c_j$ is the candidate set for $t_j$, which is an intersection of the synonym set and the stolen red list. 
In greedy search-based strategy, we sort tokens in $T^c_j$ in descending order of similarity and 
select  the most similar token to replace tokens in $T^s_g$.
This method effectively removes the watermark from the text; however, this may affect text quality (c.f., Table~\ref{tab:perplexity}). 

\subsection{Gumbel Softmax}

To mitigate the impact of watermark removal on text quality, we try to select proper substitutions by optimizing the perplexity through gradient descent. 
However, due to tokens being discrete, directly optimizing the embedding is not feasible. 
The Gumbel Softmax \cite{jang2016categorical} method can avoid the issue of non-differentiability in gradient optimization due to discrete data through sampling.
Therefore, we propose to employ Gumbel Softmax-based watermark removal to find appropriate substitutions. 

Initially, the candidate tokens in $T^c_j$ are transformed into embeddings using a tokenizer. Then, these embeddings are  organized into a matrix $E^t_j$, where each row represents an embedding of the tokens in $T^c_j$. Using the Gumbel Softmax method, one-hot vectors are sampled, and the outer product of these one-hot vectors with $E^t_j$ yields the selected vector $e'_j$. 
Lastly, we select appropriate tokens by optimizing perplexity with gradient descent:
\begin{IEEEeqnarray}{l}
    e'_j=\mathrm{softmax}(x+\varepsilon) \times E^t_j,~\varepsilon \sim \mathrm{Gumbel}(0,1), \\
    E_i=\{e'_j\}, t_j \in S_i,\\
    \min \mathrm{PPL}(E_i),
\end{IEEEeqnarray}
where 
$\mathrm{PPL}$ is the function of perplexity. We follow the same way in \cite{kirchenbauer2023watermark} and define perplexity as the exponentiated average negative log-likelihood of a sentence. 

\begin{table}[t]
  \centering
  \caption{AS1 attacker performance of watermark removal against LLaMA-2-7B.
  $G_{avg}^b$ and $G_{avg}^a$ are average number of green tokens before and after removal, GRR$=G_{avg}^a/G_{avg}^b$ is the rate of remaining green tokens.}
  \resizebox{0.48\textwidth}{!}{
    \begin{tabular}{cr|r|rr|rr|rr|rr}
    \toprule
          &       &       & \multicolumn{2}{c|}{Vanilla} & \multicolumn{2}{c|}{Oracle} & \multicolumn{2}{c|}{Pro} & \multicolumn{2}{c}{Freq.} \\
    \multicolumn{1}{r}{\makecell{Watermark\\Setting}} & \makecell{Dataset\\Size} & \multicolumn{1}{c|}{$G_{avg}^b$} & \multicolumn{1}{c}{$G_{avg}^a(\downarrow)$} & \multicolumn{1}{c|}{GRR$(\downarrow)$} & \multicolumn{1}{c}{$G_{avg}^a(\downarrow)$} & \multicolumn{1}{c|}{GRR$(\downarrow)$} & \multicolumn{1}{c}{$G_{avg}^a(\downarrow)$} & \multicolumn{1}{c|}{GRR$(\downarrow)$} & \multicolumn{1}{c}{$G_{avg}^a(\downarrow)$} & \multicolumn{1}{c}{GRR$(\downarrow)$} \\
    \midrule
    \multirow{4}[2]{*}{\makecell{$\gamma=0.25$\\$\delta=2$}} & 4000  & 73.66 & 28.88 & 39.21$\%$ & 5.80  & 7.88$\%$ & 21.03 & 28.55$\%$ & 38.72 & 52.56$\%$ \\
          & 10000 & 73.66 & 22.01 & 29.87$\%$ & 4.28  & 5.81$\%$ & 15.61 & 21.19$\%$ & 37.45 & 50.84$\%$ \\
          & 20000 & 73.66 & 15.47 & 21.00$\%$ & 3.98  & 5.41$\%$ & 15.51 & 21.05$\%$ & 37.10 & 50.37$\%$ \\
          & 40000 & 73.66 & 15.43 & 20.95$\%$ & 3.68  & 4.99$\%$ & 9.90  & 13.44$\%$ & 37.13 & 50.41$\%$ \\
    \midrule
    \multirow{4}[2]{*}{\makecell{$\gamma=0.25$\\$\delta=4$}} & 4000  & 69.73 & 27.90 & 40.02$\%$ & 3.55  & 5.09$\%$ & 21.69 & 31.11$\%$ & 33.34 & 47.81$\%$ \\
          & 10000 & 69.73 & 22.59 & 32.40$\%$ & 2.85  & 4.09$\%$ & 20.51 & 29.42$\%$ & 33.11 & 47.49$\%$ \\
          & 20000 & 69.73 & 21.17 & 30.36$\%$ & 2.71  & 3.88$\%$ & 20.46 & 29.34$\%$ & 33.71 & 48.35$\%$ \\
          & 40000 & 69.73 & 16.94 & 24.29$\%$ & 2.41  & 3.46$\%$ & 20.20 & 28.97$\%$ & 34.04 & 48.81$\%$ \\
    \midrule
    \multirow{4}[2]{*}{\makecell{$\gamma=0.5$\\$\delta=2$}} & 4000  & 115.86 & 47.86 & 41.31$\%$ & 16.39 & 14.15$\%$ & 42.04 & 36.29$\%$ & 81.24 & 70.12$\%$ \\
          & 10000 & 115.86 & 43.50 & 37.55$\%$ & 12.64 & 10.91$\%$ & 41.23 & 35.59$\%$ & 78.39 & 67.66$\%$ \\
          & 20000 & 115.86 & 30.78 & 26.56$\%$ & 12.23 & 10.55$\%$ & 38.39 & 33.13$\%$ & 77.71 & 67.08$\%$ \\
          & 40000 & 115.86 & 27.28 & 23.54$\%$ & 11.87 & 10.24$\%$ & 32.30 & 27.88$\%$ & 77.09 & 66.53$\%$ \\
    \midrule
    \multirow{4}[2]{*}{\makecell{$\gamma=0.5$\\$\delta=4$}} & 4000  & 115.72 & 46.76 & 40.41$\%$ & 13.22 & 11.42$\%$ & 37.62 & 32.51$\%$ & 74.38 & 64.28$\%$ \\
          & 10000 & 115.72 & 41.04 & 35.47$\%$ & 10.95 & 9.46$\%$ & 28.15 & 24.33$\%$ & 72.72 & 62.85$\%$ \\
          & 20000 & 115.72 & 35.85 & 30.98$\%$ & 10.24 & 8.85$\%$ & 28.29 & 24.45$\%$ & 72.94 & 63.03$\%$ \\
          & 40000 & 115.72 & 28.99 & 25.05$\%$ & 9.82  & 8.48$\%$ & 27.44 & 23.72$\%$ & 72.38 & 62.55$\%$ \\
    \bottomrule
    \end{tabular}%
  \label{tab:remove_as1_llama}%
  }
  \vspace{-0.5cm}
\end{table}%

\section{Experiments}
\subsection{Experimental Settings}
\paragraph{Models and Datasets} 
We follow the experiment settings in previous studies \cite{kirchenbauer2023reliability, kirchenbauer2023watermark, zhao2023provable}. 
The LLMs used in the experiments are OPT-1.3B \cite{zhang2022opt} and LLaMA-2-7B \cite{touvron2023llama2}. 
We randomly sample text from the C4 dataset~\cite{2019t5} as prompts to query the LLM for generating watermarked text. These watermarked texts, along with an equal number of natural texts, constitute the experimental datasets.  
The dataset sizes are 4000, 10000, 20000, and 40000, respectively. 
Gurobi \cite{gurobi} is utilized as the solver for the mixed integer programming. 

\paragraph{Parameter Setting} The two primary parameters for injecting watermark into text generated by LLM are the size of the green list $\gamma$ and the perturbation $\delta$ applied to green tokens.
In line with the research settings of prior work~\cite{kirchenbauer2023reliability, kirchenbauer2023watermark, zhao2023provable}, we establish 4 scenarios: ($\gamma=0.25, \delta=2$), ($\gamma=0.25, \delta=4$), ($\gamma=0.5, \delta=2$), ($\gamma=0.5, \delta=4$). 
The watermark detection threshold $z^*$ is set as 4. 
In all experiments, the values of $\hat{\beta}$ and $\tilde{\beta}$ range from $[0.9, 1]$. 

\paragraph{Metrics}
A desirable stealing method should acquire a large green list with a high true positive rate. Thus, each attack method is evaluated based on three crucial dimensions: 
(1) the number of tokens in the stolen green list ($N_g$); 
(2) the number of true green tokens in the stolen green list ($N_t$); 
(3) $\mathrm{Precision} (\uparrow) = N_t/N_g$. 
It is worth noting that our assessment of stealing performance is primarily based on $\mathrm{Precision}$ rather than Recall, as achieving high Recall without maintaining $\mathrm{Precision}$ can negatively impact effective watermark removal.
For example, the most extreme attack would involve marking all tokens in the entire vocabulary as green and then stealing them. While this approach would achieve 100\% recall, the high false positive rate would severely hinder watermark removal, as the attacker would need to replace or remove all tokens marked as green.

We also use the ratio of the average number of green tokens before removal to the average number of green tokens remaining after removal, denoted as Green token Remaining Rate (GRR)$(\downarrow)$, to assess the attacker's capability in watermark removal. 

\paragraph{Baseline Method} 
In this work, we compare our method with the frequency-based green list stealing approach, where tokens are categorized as green if their frequency is higher in the watermark dataset than in the natural dataset~\cite{zhao2023provable}. 
To the best of our knowledge, this is the only existing method for stealing the green list from LLM watermarks.


\begin{table}[t]
  \centering
  \caption{
  AS2 attacker performance of green list stealing against LLaMA-2-7B. 
  }
  \resizebox{0.43\textwidth}{!}{
    \begin{tabular}{cr|rrr|rrr}
    \toprule
          \multirow{2}[2]{*}{\makecell{Watermark\\Setting}} & \multicolumn{1}{c|}{\multirow{2}[2]{*}{\makecell{Dataset\\Size}}}& \multicolumn{3}{c|}{Ours} & \multicolumn{3}{c}{Freq.} \\
          &       & \multicolumn{1}{c}{$N_g$} & \multicolumn{1}{c}{$N_t$} & \multicolumn{1}{c|}{$\mathrm{Precision}(\uparrow)$} & \multicolumn{1}{c}{$N_g$} & \multicolumn{1}{c}{$N_t$} & \multicolumn{1}{c}{$\mathrm{Precision}(\uparrow)$} \\
    \midrule
    \multirow{4}[2]{*}{\makecell{$\gamma=0.25$\\$\delta=2$}} & 4000  & 3165  & 2003  & 63.29$\%$ & 6032  & 2782  & 46.12$\%$ \\
          & 10000 & 2852  & 2069  & 72.55$\%$ & 6613  & 3223  & 48.74$\%$ \\
          & 20000 & 2582  & 2056  & 79.63$\%$ & 6727  & 3505  & 52.10$\%$ \\
          & 40000 & 2393  & 1990  & 83.16$\%$ & 6680  & 3693  & 55.28$\%$ \\
    \midrule
    \multirow{4}[2]{*}{\makecell{$\gamma=0.25$\\$\delta=4$}} & 4000  & 3884  & 2813  & 72.43$\%$ & 4392  & 2882  & 65.62$\%$ \\
          & 10000 & 4466  & 3347  & 74.94$\%$ & 4736  & 3275  & 69.15$\%$ \\
          & 20000 & 4443  & 3481  & 78.35$\%$ & 4937  & 3517  & 71.24$\%$ \\
          & 40000 & 4969  & 3923  & 78.95$\%$ & 5062  & 3754  & 74.16$\%$ \\
    \midrule
    \multirow{4}[2]{*}{\makecell{$\gamma=0.5$\\$\delta=2$}} & 4000  & 6712  & 5149  & 76.71$\%$ & 6881  & 5080  & 73.83$\%$ \\
          & 10000 & 6864  & 5569  & 81.13$\%$ & 7938  & 6054  & 76.27$\%$ \\
          & 20000 & 7029  & 5872  & 83.54$\%$ & 8510  & 6616  & 77.74$\%$ \\
          & 40000 & 7902  & 6677  & 84.50$\%$ & 8828  & 7028  & 79.61$\%$ \\
    \midrule
    \multirow{4}[2]{*}{\makecell{$\gamma=0.5$\\$\delta=4$}} & 4000  & 6095  & 5256  & 86.23$\%$ & 6284  & 5249  & 83.53$\%$ \\
          & 10000 & 6868  & 6056  & 88.18$\%$ & 7386  & 6275  & 84.96$\%$ \\
          & 20000 & 6296  & 5749  & 91.31$\%$ & 7918  & 6839  & 86.37$\%$ \\
          & 40000 & 8511  & 7668  & 90.10$\%$ & 8253  & 7265  & 88.03$\%$ \\
    \bottomrule
    \end{tabular}%
  \label{tab:AS2_steal_llama}%
  }
\end{table}%
\begin{table}[]
  \centering
  \caption{
    AS2 attacker performance of watermark removal using greedy search  against OPT-1.3B and LLaMA-2-7B. 
    Ours and Freq. refer to our method and the frequency-based method, respectively. $G_{avg}^b$ and $G_{avg}^a$ represent the average number of green tokens in each sentence before and after the watermark removal.
  }
  \resizebox{0.48\textwidth}{!}{
    \begin{tabular}{cr|r|rr|rr|r|rr|rr}
    \toprule
          &       & \multicolumn{5}{c|}{OPT}              & \multicolumn{5}{c}{LLaMA} \\
    \multirow{2}[1]{*}{\makecell{Watermark\\Setting}} & \multicolumn{1}{c|}{\multirow{2}[1]{*}{\makecell{Dataset\\Size}}} & \multicolumn{1}{c|}{\multirow{2}[1]{*}{$G_{avg}^b$}} & \multicolumn{2}{c|}{$G_{avg}^a(\downarrow)$} & \multicolumn{2}{c|}{GRR$(\downarrow)$} & \multicolumn{1}{c|}{\multirow{2}[1]{*}{$G_{avg}^b$}} & \multicolumn{2}{c|}{$G_{avg}^a(\downarrow)$} & \multicolumn{2}{c}{GRR$(\downarrow)$} \\
          &       &       & \multicolumn{1}{c}{Ours} & \multicolumn{1}{c|}{Freq.} & \multicolumn{1}{c}{Ours} & \multicolumn{1}{c|}{Freq.} &       & \multicolumn{1}{c}{Ours} & \multicolumn{1}{c|}{Freq.} & \multicolumn{1}{c}{Ours} & \multicolumn{1}{c}{Freq.} \\
    \midrule
    \multirow{4}[2]{*}{\makecell{$\gamma=0.25$\\$\delta=2$}} & 4000  & 67.31 & 11.53 & 21.16 & 17.13$\%$ & 31.43$\%$ & 71.17 & 10.38 & 36.62 & 14.58$\%$ & 51.46$\%$ \\
          & 10000 & 67.31 & 10.42 & 19.24 & 15.48$\%$ & 28.59$\%$ & 71.17 & 9.62  & 35.84 & 13.52$\%$ & 50.35$\%$ \\
          & 20000 & 67.31 & 7.42  & 18.42 & 11.02$\%$ & 27.37$\%$ & 71.17 & 9.53  & 35.10 & 13.40$\%$ & 49.32$\%$ \\
          & 40000 & 67.31 & 7.75  & 17.92 & 11.51$\%$ & 26.63$\%$ & 71.17 & 9.64  & 34.90 & 13.55$\%$ & 49.04$\%$ \\
    \midrule
    \multirow{4}[2]{*}{\makecell{$\gamma=0.25$\\$\delta=4$}} & 4000  & 51.17 & 17.17 & 14.83 & 33.55$\%$ & 28.98$\%$ & 71.13 & 8.32  & 34.36 & 11.70$\%$ & 48.30$\%$ \\
          & 10000 & 51.17 & 7.57  & 13.56 & 14.80$\%$ & 26.51$\%$ & 71.13 & 7.45  & 34.09 & 10.47$\%$ & 47.92$\%$ \\
          & 20000 & 51.17 & 6.69  & 13.11 & 13.07$\%$ & 25.62$\%$ & 71.13 & 7.38  & 34.63 & 10.38$\%$ & 48.68$\%$ \\
          & 40000 & 51.17 & 5.67  & 12.87 & 11.09$\%$ & 25.15$\%$ & 71.13 & 7.58  & 34.88 & 10.66$\%$ & 49.04$\%$ \\
    \midrule
    \multirow{4}[2]{*}{\makecell{$\gamma=0.5$\\$\delta=2$}} & 4000  & 122.62 & 32.06 & 50.99 & 26.14$\%$ & 41.58$\%$ & 122.08 & 31.06 & 83.10 & 25.44$\%$ & 68.07$\%$ \\
          & 10000 & 122.62 & 27.92 & 46.68 & 22.77$\%$ & 38.06$\%$ & 122.08 & 29.53 & 80.53 & 24.19$\%$ & 65.96$\%$ \\
          & 20000 & 122.62 & 41.57 & 44.88 & 33.90$\%$ & 36.60$\%$ & 122.08 & 33.14 & 79.40 & 27.14$\%$ & 65.04$\%$ \\
          & 40000 & 122.62 & 40.84 & 43.41 & 33.31$\%$ & 35.40$\%$ & 122.08 & 31.99 & 79.13 & 26.21$\%$ & 64.82$\%$ \\
    \midrule
    \multirow{4}[2]{*}{\makecell{$\gamma=0.5$\\$\delta=4$}} & 4000  & 122.98 & 41.03 & 47.00 & 33.36$\%$ & 38.22$\%$ & 115.97 & 25.52 & 75.03 & 22.01$\%$ & 64.70$\%$ \\
          & 10000 & 122.98 & 37.52 & 43.04 & 30.51$\%$ & 35.00$\%$ & 115.97 & 27.43 & 73.41 & 23.65$\%$ & 63.30$\%$ \\
          & 20000 & 122.98 & 34.69 & 41.17 & 28.21$\%$ & 33.48$\%$ & 115.97 & 30.46 & 73.18 & 26.26$\%$ & 63.10$\%$ \\
          & 40000 & 122.98 & 31.97 & 39.94 & 26.00$\%$ & 32.47$\%$ & 115.97 & 20.34 & 72.58 & 17.54$\%$ & 62.59$\%$ \\
    \bottomrule
    \end{tabular}%
  \label{tab:remove}%
  }
  \vspace{-0.5cm}
\end{table}%

\begin{table}[t]
  \centering
  \caption{
     AS2 attacker performance of stealing green list when handling the dataset with different proportion of erroneous samples ($r_c$); the LLM is OPT-1.3B, and $\gamma=0.25, \delta=2$. 
  }
  \resizebox{0.42\textwidth}{!}{
    \begin{tabular}{rr|ccc|ccc}
    \toprule
    \multirow{2}[2]{*}{\makecell{Dataset\\Size}} & & \multicolumn{3}{c|}{Ours} & \multicolumn{3}{c}{Freq.} \\
     &$r_c$& $N_g$ & $N_t$ & $\mathrm{Precision}(\uparrow)$  & $N_g$ & $N_t$ & $\mathrm{Precision}(\uparrow)$ \\
    \midrule
    10000  & 0.1   & 3755  & 2923  & 77.84$\%$ & 13842 & 6602  & 47.70$\%$ \\
    10000  & 0.3   & 5639  & 2767  & 49.07$\%$ & 15305 & 5846  & 38.20$\%$ \\
    20000 & 0.5   & 7478  & 4598  & 61.49$\%$ & 20367 & 5444  & 26.73$\%$ \\
    20000 & 0.7   & 5851  & 3403  & 58.16$\%$ & 18942 & 3078  & 16.25$\%$ \\
    \bottomrule
    \end{tabular}%
  \label{tab:lamda_steal}%
  }
  
\end{table}%

\begin{table}[t]
  \centering
  \caption{
    AS2 attacker performance of watermark removal using
greedy search when handling the dataset with different proportions of erroneous samples; the LLM is OPT-1.3B, and $\gamma=0.25, \delta=2$. 
    }
  \resizebox{0.33\textwidth}{!}{
    \begin{tabular}{rr|c|cc|cc}
    \toprule
    \multicolumn{1}{c}{\multirow{2}[2]{*}{\makecell{Dataset\\Size}}} &       &       & \multicolumn{2}{c|}{Ours} & \multicolumn{2}{c}{Freq.} \\
          & \multicolumn{1}{c|}{$r_c$} & $G_{avg}^b$ & $G_{avg}^a(\downarrow)$ & GRR$(\downarrow)$   & $G_{avg}^a(\downarrow)$ & GRR$(\downarrow)$ \\
    \midrule
    10000 & 0.1   & 67.31 & 9.83  & 14.60$\%$ & 19.97 & 29.67$\%$ \\
    10000 & 0.3   & 67.31 & 14.53 & 21.59$\%$ & 22.67 & 33.69$\%$ \\
    20000 & 0.5   & 67.31 & 38.63 & 57.39$\%$ & 45.67 & 67.86$\%$ \\
    20000 & 0.7   & 67.31 & 42.06 & 62.49$\%$ & 82.98 & 123.29$\%$ \\
    \bottomrule
    \end{tabular}%
  \label{tab:lambda_remove}%
  }
\end{table}%

\begin{table*}[htbp]
  \centering
  \caption{AS2 attacker performance of 3-key green list stealing against OPT-1.3B and LLaMA-2-7B. }
  \resizebox{0.95\textwidth}{!}{
    \begin{tabular}{rrr|rrr|rrr|rrr|rrr|rrr|rrr}
    \toprule
          &       &       & \multicolumn{6}{c|}{Green List 1}                    & \multicolumn{6}{c|}{Green List 2}                    & \multicolumn{6}{c}{Green List 3} \\
          &       & \multirow{2}[2]{*}{\makecell{Dataset\\Size}} & \multicolumn{3}{c|}{Ours} & \multicolumn{3}{c|}{Freq.} & \multicolumn{3}{c|}{Ours} & \multicolumn{3}{c|}{Freq.} & \multicolumn{3}{c|}{Ours} & \multicolumn{3}{c}{Freq.} \\
    \multicolumn{1}{l}{Model} & \multicolumn{1}{c}{$\gamma$} &   & \multicolumn{1}{c}{$N_g$} & \multicolumn{1}{c}{$N_t$} & \multicolumn{1}{c|}{$\mathrm{Precision}(\uparrow)$} & \multicolumn{1}{c}{$N_g$} & \multicolumn{1}{c}{$N_t$} & \multicolumn{1}{c|}{$\mathrm{Precision}(\uparrow)$} & \multicolumn{1}{c}{$N_g$} & \multicolumn{1}{c}{$N_t$} & \multicolumn{1}{c|}{$\mathrm{Precision}(\uparrow)$} & \multicolumn{1}{c}{$N_g$} & \multicolumn{1}{c}{$N_t$} & \multicolumn{1}{c|}{$\mathrm{Precision}(\uparrow)$} & \multicolumn{1}{c}{$N_g$} & \multicolumn{1}{c}{$N_t$} & \multicolumn{1}{c|}{$\mathrm{Precision}(\uparrow)$} & \multicolumn{1}{c}{$N_g$} & \multicolumn{1}{c}{$N_t$} & \multicolumn{1}{c}{$\mathrm{Precision}(\uparrow)$} \\
    \midrule
    LLaMA & 0.25  & 6000  & 2154  & 1383  & 0.6421 & 2000  & 821   & 0.4105 & 2141  & 1344  & 0.6277 & 2000  & 804   & 0.4020 & 2063  & 1302  & 0.6311 & 2000  & 796   & 0.3980 \\
   LLaMA & 0.25  & 12000 & 1995  & 1513  & 0.7584 & 2000  & 836   & 0.4180 & 1995  & 1455  & 0.7293 & 2000  & 829   & 0.4145 & 1999  & 1418  & 0.7094 & 2000  & 810   & 0.4050 \\
    LLaMA & 0.5   & 6000  & 2152  & 1946  & 0.9043 & 2000  & 1412  & 0.7060 & 2263  & 1935  & 0.8551 & 2000  & 1333  & 0.6665 & 2257  & 1737  & 0.7696 & 2000  & 1148  & 0.5740 \\
    LLaMA & 0.5   & 12000 & 1998  & 1825  & 0.9134 & 2000  & 1433  & 0.7165 & 2002  & 1821  & 0.9096 & 2000  & 1334  & 0.6670 & 1997  & 1713  & 0.8578 & 2000  & 1151  & 0.5755 \\
    OPT   & 0.25  & 6000  & 3007  & 1957  & 0.6508 & 3000  & 1300  & 0.4333 & 3003  & 1918  & 0.6387 & 3000  & 1296  & 0.4320 & 2992  & 1959  & 0.6547 & 3000  & 1171  & 0.3903 \\
    OPT   & 0.5   & 6000  & 2995  & 2549  & 0.8511 & 3000  & 1954  & 0.6513 & 2997  & 2538  & 0.8468 & 3000  & 1888  & 0.6293 & 2996  & 2565  & 0.8561 & 3000  & 1886  & 0.6287 \\
    \bottomrule
    \end{tabular}%
  \label{tab:multi_steal}%
  }
\end{table*}%

\begin{table*}[]
  \centering
  \caption{AS2 attacker performance of removal for 3-key watermark against OPT-1.3B and LLaMA-2-7B. }
  \resizebox{0.9\textwidth}{!}{
    \begin{tabular}{rrr|r|rr|ll|r|rr|ll|r|rr|ll}
    \toprule
          &       &       & \multicolumn{5}{c|}{Green List 1}     & \multicolumn{5}{c|}{Green List 2}     & \multicolumn{5}{c}{Green List 3} \\
          &       & \multicolumn{1}{c|}{\multirow{2}[1]{*}{\makecell{Dataset\\Size}}} & \multicolumn{1}{c|}{\multirow{2}[1]{*}{$G_{avg}^b$}} & \multicolumn{2}{c|}{$G_{avg}^a(\downarrow)$} & \multicolumn{2}{c|}{GRR$(\downarrow)$} & \multicolumn{1}{c|}{\multirow{2}[1]{*}{$G_{avg}^b$}} & \multicolumn{2}{c|}{$G_{avg}^a(\downarrow)$} & \multicolumn{2}{c|}{GRR$(\downarrow)$} & \multicolumn{1}{c|}{\multirow{2}[1]{*}{$G_{avg}^b$}} & \multicolumn{2}{c|}{$G_{avg}^a(\downarrow)$} & \multicolumn{2}{c}{GRR$(\downarrow)$} \\
    \multicolumn{1}{c}{Model} & \multicolumn{1}{c}{$\gamma$} &       &       & \multicolumn{1}{c}{Ours} & \multicolumn{1}{c|}{Freq.} & \multicolumn{1}{c}{Ours} & \multicolumn{1}{c|}{Freq.} &       & \multicolumn{1}{c}{Ours} & \multicolumn{1}{c|}{Freq.} & \multicolumn{1}{c}{Ours} & \multicolumn{1}{c|}{Freq.} &       & \multicolumn{1}{c}{Ours} & \multicolumn{1}{c|}{Freq.} & \multicolumn{1}{c}{Ours} & \multicolumn{1}{c}{Freq.} \\
    \midrule
    LLaMA & 0.25  & 6000  & 77.75 & 47.19 & 72.19 & 60.68$\%$ & 92.84$\%$ & 69.77 & 38.95 & 62.59 & 55.82$\%$ & 89.71$\%$ & 75.55 & 35.42 & 67.69 & 46.88$\%$ & 89.59$\%$ \\
    LLaMA & 0.25  & 12000 & 77.75 & 45.93 & 73.60 & 59.07$\%$ & 94.66$\%$ & 69.77 & 39.15 & 64.66 & 56.11$\%$ & 92.67$\%$ & 75.55 & 35.61 & 69.79 & 47.13$\%$ & 92.38$\%$ \\
    LLaMA & 0.5   & 6000  & 121.75 & 86.20 & 118.36 & 70.80$\%$ & 97.21$\%$ & 130.12 & 95.09 & 126.01 & 73.08$\%$ & 96.84$\%$ & 99.87 & 78.16 & 97.90 & 78.26$\%$ & 98.02$\%$ \\
    LLaMA & 0.5   & 12000 & 121.75 & 91.58 & 119.31 & 75.22$\%$ & 97.99$\%$ & 130.12 & 90.80 & 127.10 & 69.79$\%$ & 97.69$\%$ & 99.87 & 74.35 & 98.28 & 74.45$\%$ & 98.40$\%$ \\
    OPT   & 0.25  & 6000  & 80.04 & 44.72 & 75.09 & 55.87$\%$ & 93.82$\%$ & 78.87 & 43.27 & 75.45 & 54.86$\%$ & 95.67$\%$ & 75.15 & 40.09 & 71.21 & 53.35$\%$ & 94.76$\%$ \\
    OPT   & 0.5   & 6000  & 117.40 & 83.45 & 115.92 & 71.08$\%$ & 98.74$\%$ & 117.14 & 82.46 & 115.08 & 70.39$\%$ & 98.24$\%$ & 117.56 & 79.45 & 115.69 & 67.58$\%$ & 98.41$\%$ \\
    \bottomrule
    \end{tabular}%
  \label{tab:multi_remove}%
  }
\end{table*}%

\begin{table}[htbp]
  \centering
  \caption{
    A comparison of perplexity between watermark removal methods based on greedy search and Gumbel Softmax. The Gumbel Softmax-based removal method achieves lower perplexity than the greedy search-based method. Raw means the perplexity of the original sentence.
  }
  \resizebox{0.47\textwidth}{!}{
    \begin{tabular}{cr|ccc|ccc}
    \toprule
          \multirow{2}[2]{*}{\makecell{Watermark\\Setting}} & \multicolumn{1}{c|}{\multirow{2}[2]{*}{\makecell{Dataset\\Size}}}& \multicolumn{3}{c|}{OPT} & \multicolumn{3}{c}{LLaMA} \\
          &       & \multicolumn{1}{c}{Raw} & \multicolumn{1}{c}{Greedy} & \multicolumn{1}{c|}{Gumbel Softmax} & \multicolumn{1}{c}{Raw} & \multicolumn{1}{c}{Greedy} & \multicolumn{1}{c}{Gumbel Softmax} \\
    \midrule
    \multirow{4}[2]{*}{\makecell{$\gamma=0.25$\\$\delta=2$}} & 4000  & 3.28  & 7.25  & 6.51  & 3.05  & 7.25  & 7.00 \\
          & 10000  & 3.28  & 7.27  & 6.52  & 3.05  & 7.27  & 7.01 \\
          & 20000 & 3.28  & 7.38  & 6.66  & 3.05  & 7.23  & 6.95 \\
          & 40000 & 3.28  & 7.35  & 6.62  & 3.05  & 7.20  & 6.94 \\
    \midrule
    \multirow{4}[2]{*}{\makecell{$\gamma=0.25$\\$\delta=4$}} & 4000  & 3.70  & 7.11  & 6.57  & 3.48  & 7.41  & 7.20 \\
          & 10000  & 3.70  & 7.76  & 7.09  & 3.48  & 7.42  & 7.23 \\
          & 20000 & 3.70  & 7.81  & 7.14  & 3.48  & 7.45  & 7.19 \\
          & 40000 & 3.70  & 7.82  & 7.14  & 3.48  & 7.42  & 7.16 \\
    \midrule
    \multirow{4}[2]{*}{\makecell{$\gamma=0.5$\\$\delta=2$}} & 4000  & 3.14  & 8.10  & 7.69  & 2.98  & 8.09  & 7.85 \\
          & 10000  & 3.14  & 8.13  & 7.83  & 2.98  & 8.01  & 7.76 \\
          & 20000 & 3.14  & 7.69  & 7.49  & 2.98  & 7.90  & 7.62 \\
          & 40000 & 3.14  & 7.62  & 7.50  & 2.98  & 7.88  & 7.59 \\
    \midrule
    \multirow{4}[2]{*}{\makecell{$\gamma=0.5$\\$\delta=4$}} & 4000  & 3.27  & 7.88  & 7.77  & 3.20  & 8.23  & 7.96 \\
          & 10000  & 3.27  & 7.87  & 7.76  & 3.20  & 8.15  & 7.86 \\
          & 20000 & 3.27  & 7.90  & 7.79  & 3.20  & 8.06  & 7.75 \\
          & 40000 & 3.27  & 7.91  & 7.79  & 3.20  & 8.19  & 7.92 \\
    \bottomrule
    \end{tabular}%
  \label{tab:perplexity}%
  }
  \vspace{-0.5cm}
\end{table}%

\subsection{AS1 Attacker Performance}
\label{sec:exp_as1}

We present the performance of AS1 attacker against LLaMA in stealing the green list in Table \ref{tab:steal_as1_llama} and the performance of watermark removal in Table \ref{tab:remove_as1_llama} for LLaMA. Additional results for OPT are provided in Appendix~\ref{app:exp} in Table~\ref{tab:steal_as1_opt} and Table~\ref{tab:remove_as1_opt}.

In terms of stealing green lists, overall, our methods (i.e., Vanilla-AS1, Oracle-AS1, Pro-AS1) exhibit stronger stealing capabilities compared to the baseline approach. For instance, when the dataset size is 40000 and the setting is $\gamma=0.25, \delta=2$, Vanilla-AS1, Oracle-AS1, and Pro-AS1 achieve $\mathrm{Precision}$ improvements of $11.14\%$, $35.83\%$, and $27.79\%$, respectively, over the frequency-based method. 
The high false positive rate of frequency-based methods arises from the presence of low-entropy tokens and numerous sentences containing green tokens near the watermark detection threshold. This results in an ambiguous and inaccurate space for identifying true green tokens. 
In contrast, our methods can identify an accurate feasible domain using explicitly defined constraints based on the watermark rules, thereby achieving higher precision. 
Oracle-AS1 achieves higher $\mathrm{Precision}$ compared to Vanilla-AS1 while stealing more true green tokens. Vanilla-AS1 typically achieves a $\mathrm{Precision}$ rate of only $70\%$, whereas Oracle-AS1 frequently surpasses $90\%$ and, in certain instances, nears $100\%$. This highlights the effectiveness of incorporating the ground truth number of green tokens $g^o$ in the optimization process. 
In general, Pro-AS1 showcases a superior approximation of Oracle-AS1's stealing performance, with a $\mathrm{Precision}$ gap of only $8.49\%$, as opposed to $5.91\%$ observed between Vanilla-AS1 and Oracle-AS1 for LLaMA. This is attributed to the Stage 1 optimization strategy (Eq.~(\ref{eq:find_boundary_abs}) ) of Pro-AS1, which identifies tighter bounds to regulate the optimization process.

Table~\ref{tab:remove_as1_llama} shows the AS1 attacker's performance in  watermark removal. In the table, we use $G_{avg}^b$ to denote the average number of green tokens before removal, and  $G_{avg}^a$ to denote the average number of green tokens after removal. Experimental results indicate that Vanilla-AS1, Oracle-AS1, and Pro-AS1 are more effective than the baseline method in removing green tokens from sentences. 
On LLaMA, the average reduction amounts to $26.36\%$, $49.75\%$, and $29.99\%$. This is attributed to the ability of mixed integer
programming-based methods to more accurately steal the true green list, thus enabling precise substitution of green tokens with red tokens during watermark removal. Conversely, frequency-based methods tend to pilfer green lists with lower $\mathrm{Precision}$, potentially leading to the erroneous replacement of red tokens within sentences. 
The results of the watermark removal demonstrate that stealing green tokens with higher $\mathrm{Precision}$ can lead to a more effective attack. 
Additional results for OPT can be found in Appendix~\ref{app:exp} in Table~\ref{tab:remove_as1_opt}.

\subsection{AS2 Attacker Performance}
\label{sec:exp_AS2}

The performance of our AS2 attacker in green list stealing for LLaMA is shown in Table~\ref{tab:AS2_steal_llama}. The results for OPT can be found in Appendix~\ref{app:exp} (Table~\ref{tab:AS2_steal_opt}). 
Since the attacker does not have access to the detector's API, they are unable to verify whether each sentence contains a watermark, leading to the presence of erroneous samples in the dataset.
In our first set of experiments, {$r_c$}, the proportion of the erroneous samples in a dataset, is set to be lower than $0.05$.  We set $p_u$ and $p_l$  to $0.99$ and $0.98$, respectively. 
Overall, our method consistently outperforms the frequency-based method in terms of $\mathrm{Precision}$ across all settings. For LLaMA, the Precision improvement ranges from  $2.07\%$ to $27.87\%$. Additionally, as the dataset size increases, the $\mathrm{Precision}$ of our method improves correspondingly, whereas the $\mathrm{Precision}$ of the frequency-based method remains relatively unchanged.
This indicates that our method exhibit stronger attack capability if a large amount of data can be collected. In fact, both natural and watermarked texts are relatively easy to obtain.

Table~\ref{tab:remove} demonstrates the performance of AS2 attacker in watermark removal using greedy search-based strategy. 
In the settings of OPT and LLaMA, our method removed an average of $77.38\%$ and $81.83\%$ of green tokens from the original watermarked texts, respectively, while the frequency-based method only removed $68.08\%$ and $43.02\%$. Across all 32 attack settings (4 Watermark Settings $\times$ 4 Data Sizes $\times$ 2 LLMs), our method reduced the average GRR to $20.39\%$, whereas the frequency-based method resulted in an average GRR nearly twice as high, at $44.46\%$.


\begin{figure*}[htbp]
    \centering
    \includegraphics[width=0.9\textwidth]{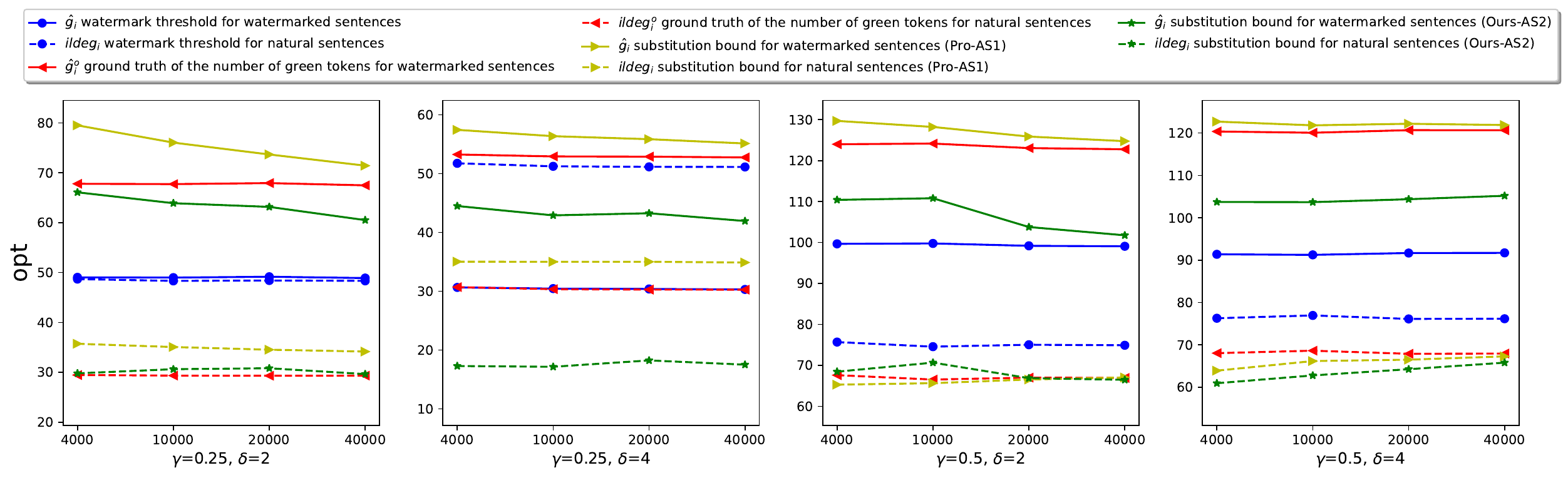}
    \includegraphics[width=0.9\textwidth]{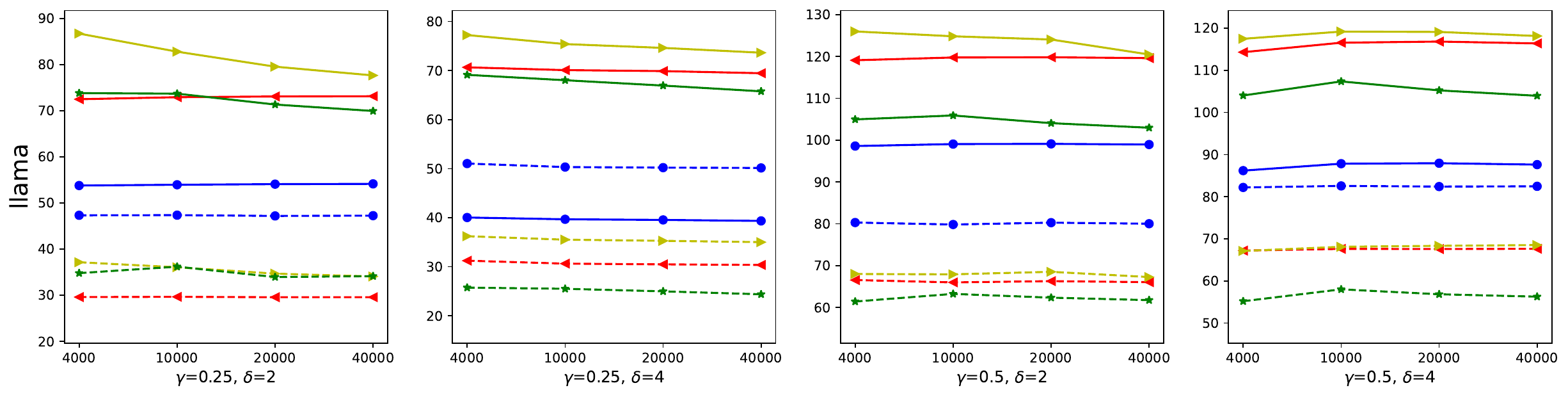}
    \caption{
    A comparison of $\hat{g}^o_i$, $g_i$, and $\hat{b}_i$ across all settings for OPT-1.3B and LLaMA-2-7B. The results show that $\hat{g}^o_i$ is consistently larger than $g_i$ in watermark text, while in natural text, $\hat{g}^o_i$ is consistently smaller than $g_i$. $\hat{b}_i$ calculated using Eq.(\ref{eq:find_boundary_abs}) is closer to $\hat{g}^o_i$ than $g_i$, and there are limited differences between $\hat{b}_i$ determined by Eq.~(\ref{eq:find_boundary_abs}) and Eq.~(\ref{eq:find_boundary_noapi}).}
    \label{fig:opt_llama_gi_bi}
    \vspace{-0.5cm}
\end{figure*}

\paragraph{The Proportion of Erroneous Samples}  
This section investigates the effect of proportion of the erroneous samples, $r_c$. 
In the following set of experiments, we set $p_u$ and $p_l$ to $0.9-r_c$ and $0.8-r_c$, respectively.
We manually construct the erroneous samples in the dataset by adding watermarked text into a natural dataset or adding natural text into a watermark dataset. 
We compare the performance of our method with frequency-based method on OPT, using $r_c=0.1, 0.3, 0.5, 0.7$ under the setting of $\gamma=0.25, \delta=2$. 

Table~\ref{tab:lamda_steal} shows the performance of stealing green list with various $r_c$. 
Our proposed method is capable of stealing the green list across all the settings, even when $r_c$ is high, while the efficacy of the frequency method drops significantly when $r_c$ raises. 
This demonstrates the effectiveness of optimizing $\lambda$ to identify the erroneous samples. 
With this capability, our method can resist countermeasures, such as the strategy of occasionally refreshing the green list to defend against frequency attacks~\cite{zhao2023provable}. 

Table~\ref{tab:lambda_remove} shows the performance of the AS2 attacker in watermark removal when handling the dataset with different proportions of the erroneous samples. 
The results show even if the proportion of erroneous sample is $70\%$, our method can remove nearly $40\%$ green tokens, and this is sufficient to evade the detection. 
However, in this circumstance, for the frequency-based method, after removal, GRR is higher than $100\%$; This is because the $\mathrm{Precision}$ of the green list stolen by the frequency-based method is too low, preventing the attacker from accurately identifying green tokens during watermark removal, resulting in many red tokens being mistakenly replaced.
We additionally use the watermark detector API to assess the sentences after watermark removal. Notably, even with the setting of $r_c = 0.7$, our AS2 attacker successfully evades the detector, with an average of $81.55\%$ of sentences being marked as not watermarked. In contrast, the frequency-based method removes the watermark from only $0.12\%$ of sentences. 


\paragraph{Multi-key Stealing} 
We now assess the performance of our attack against the  watermark schemes with multiple keys~\cite{kirchenbauer2023watermark, kirchenbauer2023reliability}. 
We set the number of keys to be 3, as the watermark with many keys can be easily removed by paraphrasing attacks~\cite{zhao2023provable, hou2023semstamp, ren2023robust}. 
Each sentence is randomly assigned a key from $K$. 
For LLaMA, we set  $\mu=2000$, and for OPT, we set $\mu=3000$ (c.f., Eq.~(\ref{eq:mu})).  
In the frequency-based method, tokens are sorted in descending order based on their frequency, and then the top-$\mu$ tokens are chosen as the green list. 
Because frequency-based method only find one green list, we report its performance by comparing with the true green lists respectively. 

Table~\ref{tab:multi_steal} shows the performance of green list stealing against the 3-key watermark scheme. 
Our method consistently achieves a stealing $\mathrm{Precision}$ that is at least $23\%$ higher than the frequency-based method, with an average $\mathrm{Precision}$ of $76.70\%$. Our mixed integer programming-based approach introduces variable $\rho^k_i$ (c.f. Eq.~(\ref{eq:sum_rho})), which accurately estimates the corresponding key to each sentence, and employs an iterative algorithm (Algorithm.~\ref{alg:multi_key}) to find the most accurate green lists that satisfy all constraints. In contrast, the baseline method only identify the token with the highest frequency among multiple green lists, leading to ineffective stealing outcomes.
Furthermore, as the dataset size increases, our method demonstrates stronger stealing capability, while the $\mathrm{Precision}$ of the frequency-based method remains at a lower level.

Table~\ref{tab:multi_remove} shows the performance of watermark removal in this setting. Our method can significantly decrease the number of the remaining green tokens by more than 30, while the frequency-based method is ineffective as the number of remaining green tokens is  unchanged.


\subsection{Estimating the Bound on the Number of Green Tokens}
In this section, we conduct experiments to analyze the effectiveness of Stage 1 in Eq.~(\ref{eq:find_boundary_abs}) and (\ref{eq:find_boundary_noapi}) in approximating $\hat{b}_i$ (or $\tilde{b}_i$) to $\hat{g}^o_i$ (or $\tilde{g}^o_i$).
As shown in Figure~\ref{fig:opt_llama_gi_bi}, the ground truth $\hat{g}^o_i$ in watermark text is always larger than the watermark detection threshold $g_i$ while $\tilde{g}^o_i$ in natural text is always smaller than $g_i$. 
This demonstrates that $\hat{g}^o_i$ and $\tilde{g}^o_i$ provides tighter bounds than $g_i$, which can benefit the optimization of mixed integer programming. 


Pro-AS1  can find $\hat{b}_i$ (c.f., Eq.~(\ref{eq:find_boundary_abs}))  that is closer to $\hat{g}^o_i$ than $g_i$. 
For LLaMA, the average differences between $\hat{b}_i$ and $\hat{g}^o_i$ for watermarked and natural sentences are only $5.18$ and $3.23$, respectively, and this difference tends to decrease as the dataset grows. Conversely, the average differences between $g_i$ and $\hat{g}^o_i$ for watermarked and natural sentences are significantly higher, at $24.64$ and $16.55$ respectively, nearly six times larger than the average differences between $\hat{b}_i$ and $\hat{g}^o_i$. A similar phenomenon can also be observed in OPT. This indicates that after Stage 1 optimization, Eq.~(\ref{eq:find_boundary_abs}) can find tighter constraints to approximate Oracle-AS1.
AS2 attackers can find $\hat{b}_i$  (c.f., Eq.~(\ref{eq:find_boundary_noapi})) that also exhibits minor difference to $\hat{g}^o_i$. 
The average difference between $\hat{b}_i$ and $\hat{g}^o_i$ are $7.31$ and $3.83$ for watermarked and natural sentences, respectively. 

\subsection{Watermark Removing With Gumbel Softmax}
 




In this section, we experimentally analyze the impact of the Gumbel Softmax-based method on text quality during watermark removal.
We adopt the victim LLMs to compute perplexity during the optimization of Gumbel Softmax-based watermark removal.  
It is worth noting that sentences with lower perplexity are more fluent than higher ones. 
As shown in  Table~\ref{tab:perplexity}, Gumbel Softmax-based methods can select appropriate tokens to remove watermarks while maintaining favorable perplexity in the results. 
Employing Gumbel Softmax to sample appropriate tokens and optimizing the perplexity of sentences with synonym replacements via gradient descent, the average perplexity of sentences generated by Gumbel Softmax-based approaches is reduced by $0.4368$ and $0.2603$ on the OPT and LLaMA models, respectively, in comparison to greedy search.


\section{Conclusion}
In this work, we have presented a novel watermark removal attack against the state-of-the-art LLM watermark scheme, employing mixed-integer programming to extract the feasible green list. 
The optimization is guided by a set of constraints derived from the watermark rules. 
Our attack can successfully steal the green list and remove the watermark even without any prior knowledge, lacking access to the watermark detector API and possessing no information about the LLMs' parameter settings or their watermark injection/detection scheme. 
Our method forms a generic framework capable of targeting both single-key and multi-key watermark schemes, including token-level and sentence-level approaches. 
This study demonstrates that the robustness of watermarks in LLMs is significantly compromised when facing stealing attacks. 

Our findings highlight the urgent need for dedicated defenses against watermark removal attacks. 
One possible direction is to incorporate synonyms of green tokens into the green list when adding watermarks, as it makes more challenging for attackers to reduce the number of green tokens in a sentence through synonym substitution. 
Although such a strategy increases the difficulty of replacing green tokens, it could reduce the diversity of the green list, making it even easier to steal. 
We also see research opportunities in developing new unbiased watermark schemes where the distribution of watermarked text maintains the same expectation as the unwatermarked distribution. 
 
\section{Acknowledgments}
This research was supported in partial by RMIT AWS Cloud Supercomputing (RACE) program and  the New Researcher Grants at Griffith. The first two authors are funded by the China Scholarship Council (CSC) from the Ministry of Education, China.

\bibliographystyle{ACM-Reference-Format}
\bibliography{ref}

\begin{table*}[ht]
  \centering
  \caption{
  AS1 attacker performance of green list stealing against OPT-1.3B. $N_g$ and $N_t$ represent the number of green tokens and the number of true green tokens, respectively. Precision = $N_t$/$N_g$.
  }
  \resizebox{0.7\textwidth}{!}{
    \begin{tabular}{cr|rrr|rrr|rrr|rrr}
    \toprule
    \multirow{2}[2]{*}{\makecell{Watermark\\Setting}} & \multicolumn{1}{c|}{\multirow{2}[2]{*}{\makecell{Dataset\\size}}} & \multicolumn{3}{c|}{Vanilla} & \multicolumn{3}{c|}{Oracle} & \multicolumn{3}{c|}{Pro} & \multicolumn{3}{c}{Frequency} \\
          &       & \multicolumn{1}{c}{$N_g$} & \multicolumn{1}{c}{$N_t$} & \multicolumn{1}{c|}{Precision$(\uparrow)$} & \multicolumn{1}{c}{$N_g$} & \multicolumn{1}{c}{$N_t$} & \multicolumn{1}{c|}{Precision$(\uparrow)$} & \multicolumn{1}{c}{$N_g$} & \multicolumn{1}{c}{$N_t$} & \multicolumn{1}{c|}{Precision$(\uparrow)$} & \multicolumn{1}{c}{$N_g$} & \multicolumn{1}{c}{$N_t$} & \multicolumn{1}{c}{Precision$(\uparrow)$} \\
    \midrule
    \multirow{4}[2]{*}{\makecell{$\gamma=0.25$\\$\delta=2$}} & 4000  & 1439  & 992   & 68.94$\%$ & 4966  & 3810  & 76.72$\%$ & 2873  & 2205  & 76.75$\%$ & 8863  & 4877  & 55.03$\%$ \\
          & 10000 & 2161  & 1688  & 78.11$\%$ & 9337  & 8711  & 93.30$\%$ & 2900  & 2477  & 85.41$\%$ & 11452 & 6406  & 55.94$\%$ \\
          & 20000 & 2958  & 2320  & 78.43$\%$ & 10577 & 10326 & 97.63$\%$ & 4188  & 3564  & 85.10$\%$ & 12567 & 7369  & 58.64$\%$ \\
          & 40000 & 3251  & 2914  & 89.63$\%$ & 11300 & 11216 & 99.26$\%$ & 3715  & 3389  & 91.22$\%$ & 12567 & 7971  & 63.43$\%$ \\
    \midrule
    \multirow{4}[2]{*}{\makecell{$\gamma=0.25$\\$\delta=4$}} & 4000  & 1678  & 1316  & 78.43$\%$ & 5465  & 4854  & 88.82$\%$ & 3387  & 2920  & 86.21$\%$ & 6691  & 5007  & 74.83$\%$ \\
          & 10000 & 1733  & 1459  & 84.19$\%$ & 9256  & 8683  & 93.81$\%$ & 3784  & 3381  & 89.35$\%$ & 8922  & 6657  & 74.61$\%$ \\
          & 20000 & 2165  & 1863  & 86.05$\%$ & 10499 & 10237 & 97.50$\%$ & 3904  & 3633  & 93.06$\%$ & 10247 & 7723  & 75.37$\%$ \\
          & 40000 & 2244  & 2076  & 92.51$\%$ & 11228 & 11146 & 99.27$\%$ & 3897  & 3704  & 95.05$\%$ & 11010 & 8535  & 77.52$\%$ \\
    \midrule
    \multirow{4}[2]{*}{\makecell{$\gamma=0.5$\\$\delta=2$}} & 4000  & 2083  & 1592  & 76.43$\%$ & 9744  & 8102  & 83.15$\%$ & 7258  & 5613  & 77.34$\%$ & 9520  & 7189  & 75.51$\%$ \\
          & 10000 & 3768  & 2972  & 78.87$\%$ & 18053 & 16855 & 93.36$\%$ & 9072  & 7273  & 80.17$\%$ & 14910 & 11084 & 74.34$\%$ \\
          & 20000 & 6233  & 4739  & 76.03$\%$ & 20882 & 20518 & 98.26$\%$ & 10307 & 8622  & 83.65$\%$ & 19283 & 14092 & 73.08$\%$ \\
          & 40000 & 7538  & 6630  & 87.95$\%$ & 22464 & 22342 & 99.46$\%$ & 11207 & 9893  & 88.28$\%$ & 23187 & 16863 & 72.73$\%$ \\
    \midrule
    \multirow{4}[2]{*}{\makecell{$\gamma=0.5$\\$\delta=4$}} & 4000  & 2189  & 1839  & 84.01$\%$ & 11033 & 9897  & 89.70$\%$ & 4401  & 3900  & 88.62$\%$ & 9305  & 7724  & 83.01$\%$ \\
          & 10000 & 3632  & 2999  & 82.57$\%$ & 18358 & 17481 & 95.22$\%$ & 5421  & 5022  & 92.64$\%$ & 14306 & 11732 & 82.01$\%$ \\
          & 20000 & 5313  & 4366  & 82.18$\%$ & 21005 & 20654 & 98.33$\%$ & 6124  & 5780  & 94.38$\%$ & 18153 & 14735 & 81.17$\%$ \\
          & 40000 & 6822  & 5797  & 84.98$\%$ & 22568 & 22450 & 99.48$\%$ & 6837  & 6496  & 95.01$\%$ & 21637 & 17538 & 81.06$\%$ \\
    \bottomrule
    \end{tabular}%
  \label{tab:steal_as1_opt}%
  \vspace{-0.3cm}
  } 
\end{table*}%
\section*{Appendix}
\appendix
The Appendix provides notations and additional experimental results for OPT-1.3B.

\section{Additional Experiment Results}
\label{app:exp}

Table~\ref{tab:steal_as1_opt} presents a comparison of the Precision of our methods and the baseline method for green list extraction on the OPT model under attack setting AS1. It is evident from the results that the Oracle-AS1 method performs the best, with the Pro method closely approximating its performance. Although the Vanilla method exhibits slightly lower Precision compared to the Oracle and Pro methods, it  outperforms the frequency-based approach. This demonstrates the superior attack capability of methods based on mixed integer programming.

Additionally, Pro-AS1 showcases a superior approximation of Oracle-AS1's stealing performance, with a Precision gap of only $6.31\%$, as opposed to $12.12\%$ observed between Vanilla-AS1 and Oracle-AS1. This indicates that the two-stage optimization significantly aids Pro-AS1 in approaching the effectiveness of Oracle-AS1, as the first-stage optimization strategy of Pro-AS1 (Eq.~(\ref{eq:find_boundary_abs})) identifies tighter bounds to regulate the optimization process.

Table~\ref{tab:remove_as1_opt} shows the experiment result of AS1 attacker performance of watermark removal against OPT-1.3B. The table reports three metrics for each method under various watermark settings: $G_{avg}^b$, $G_{avg}^a$ and GRR. 
Compared to the frequency-based method, our three methods exhibit an average reduction in GRR of $3.36\%$, $24.27\%$, and $15.49\%$, respectively.
Although all four methods result in a reduction of $G_{avg}^a$, the clear GRR trend observed in most attack settings is Oracle-AS1 < Pro-AS1 < Vanilla-AS1 < frequency-based method. This indicates a significant advantage of our proposed mixed integer programming-based method in watermark removal.

Table~\ref{tab:AS2_steal_opt} presents the capability of stealing the green list from the OPT model under attack setting AS2. Across all scenarios, our method exhibits an average Precision that is $0.1549$ higher than the frequency-based approach. 

\begin{table}[t]
  \centering
  \caption{
  AS1 attacker performance of watermark removal against OPT-1.3B.
  $G_{avg}^b$ and $G_{avg}^a$ are average number of green tokens before and after removal, GRR$=G_{avg}^a/G_{avg}^b$ is the rate of remaining green tokens.
  }
  \resizebox{0.48\textwidth}{!}{
    \begin{tabular}{cr|r|rr|rr|rr|rr}
    \toprule
          &       &       & \multicolumn{2}{c|}{Vanilla} & \multicolumn{2}{c|}{Oracle} & \multicolumn{2}{c|}{Pro} & \multicolumn{2}{c}{Freq.} \\
    \multicolumn{1}{r}{\makecell{Watermark\\Setting}} & \makecell{Dataset\\Size} & \multicolumn{1}{c|}{$G_{avg}^b$} & \multicolumn{1}{c}{$G_{avg}^a(\downarrow)$} & \multicolumn{1}{c|}{GRR$(\downarrow)$} & \multicolumn{1}{c}{$G_{avg}^a(\downarrow)$} & \multicolumn{1}{c|}{GRR$(\downarrow)$} & \multicolumn{1}{c}{$G_{avg}^a(\downarrow)$} & \multicolumn{1}{c|}{GRR$(\downarrow)$} & \multicolumn{1}{c}{$G_{avg}^a(\downarrow)$} & \multicolumn{1}{c}{GRR$(\downarrow)$} \\
    \midrule
    \multirow{4}[2]{*}{\makecell{$\gamma=0.25$\\$\delta=2$}} & 4000  & 68.01 & 21.74 & 31.96$\%$ & 7.15  & 10.51$\%$ & 11.24 & 16.53$\%$ & 21.54 & 31.67$\%$ \\
          & 10000 & 68.01 & 18.76 & 27.59$\%$ & 2.70  & 3.96$\%$ & 11.17 & 16.43$\%$ & 19.89 & 29.25$\%$ \\
          & 20000 & 68.01 & 12.83 & 18.86$\%$ & 2.22  & 3.26$\%$ & 8.19  & 12.05$\%$ & 19.27 & 28.34$\%$ \\
          & 40000 & 68.01 & 9.65  & 14.19$\%$ & 2.12  & 3.12$\%$ & 8.42  & 12.37$\%$ & 18.80 & 27.65$\%$ \\
    \midrule
    \multirow{4}[2]{*}{\makecell{$\gamma=0.25$\\$\delta=4$}} & 4000  & 52.45 & 19.59 & 37.35$\%$ & 4.81  & 9.16$\%$ & 7.12  & 13.57$\%$ & 15.02 & 28.63$\%$ \\
          & 10000 & 52.45 & 17.57 & 33.50$\%$ & 2.49  & 4.75$\%$ & 6.63  & 12.65$\%$ & 13.66 & 26.04$\%$ \\
          & 20000 & 52.45 & 15.39 & 29.35$\%$ & 2.15  & 4.10$\%$ & 6.47  & 12.33$\%$ & 13.17 & 25.10$\%$ \\
          & 40000 & 52.45 & 10.95 & 20.88$\%$ & 2.04  & 3.89$\%$ & 6.45  & 12.29$\%$ & 12.91 & 24.60$\%$ \\
    \midrule
    \multirow{4}[2]{*}{\makecell{$\gamma=0.5$\\$\delta=2$}} & 4000  & 123.19 & 45.18 & 36.68$\%$ & 19.80 & 16.07$\%$ & 21.52 & 17.47$\%$ & 49.82 & 40.44$\%$ \\
          & 10000 & 123.19 & 37.47 & 30.42$\%$ & 11.27 & 9.15$\%$ & 21.18 & 17.19$\%$ & 45.47 & 36.91$\%$ \\
          & 20000 & 123.19 & 35.64 & 28.93$\%$ & 9.59  & 7.78$\%$ & 19.67 & 15.96$\%$ & 43.47 & 35.29$\%$ \\
          & 40000 & 123.19 & 29.88 & 24.25$\%$ & 9.09  & 7.38$\%$ & 17.29 & 14.04$\%$ & 41.90 & 34.01$\%$ \\
    \midrule
    \multirow{4}[2]{*}{\makecell{$\gamma=0.5$\\$\delta=4$}} & 4000  & 120.56 & 43.54 & 36.11$\%$ & 16.96 & 14.07$\%$ & 30.62 & 25.40$\%$ & 47.06 & 39.04$\%$ \\
          & 10000 & 120.56 & 39.13 & 32.46$\%$ & 10.95 & 9.08$\%$ & 27.32 & 22.66$\%$ & 43.13 & 35.77$\%$ \\
          & 20000 & 120.56 & 34.20 & 28.37$\%$ & 9.38  & 7.78$\%$ & 24.86 & 20.62$\%$ & 41.14 & 34.12$\%$ \\
          & 40000 & 120.56 & 30.30 & 25.14$\%$ & 8.88  & 7.37$\%$ & 24.53 & 20.35$\%$ & 39.65 & 32.89$\%$ \\
    \bottomrule
    \end{tabular}%
    \label{tab:remove_as1_opt}
  }
\end{table}%




\begin{table}[t]
  \centering
  \caption{
  AS2 attacker performance of green list stealing against OPT-1.3B.  
  }
  \resizebox{0.43\textwidth}{!}{
    \begin{tabular}{cr|rrr|rrr}
    \toprule
          \multirow{2}[2]{*}{\makecell{Watermark\\Setting}} & \multicolumn{1}{c|}{\multirow{2}[2]{*}{\makecell{Dataset\\Size}}}& \multicolumn{3}{c|}{Ours} & \multicolumn{3}{c}{Freq.} \\
          &       & \multicolumn{1}{c}{$N_g$} & \multicolumn{1}{c}{$N_t$} & \multicolumn{1}{c|}{$\mathrm{Precision}(\uparrow)$} & \multicolumn{1}{c}{$N_g$} & \multicolumn{1}{c}{$N_t$} & \multicolumn{1}{c}{$\mathrm{Precision}(\uparrow)$} \\
    \midrule
    \multirow{4}[2]{*}{\makecell{$\gamma=0.25$\\$\delta=2$}} & 4000  & 3333  & 2516  & 75.49$\%$ & 10929 & 5536  & 50.65$\%$ \\
          & 10000 & 3173  & 2693  & 84.87$\%$ & 14244 & 7249  & 50.89$\%$ \\
          & 20000 & 4903  & 4109  & 83.81$\%$ & 16006 & 8360  & 52.23$\%$ \\
          & 40000 & 4286  & 3911  & 91.25$\%$ & 16511 & 9100  & 55.11$\%$ \\
    \midrule
    \multirow{4}[2]{*}{\makecell{$\gamma=0.25$\\$\delta=4$}} & 4000  & 3555  & 3095  & 87.06$\%$ & 7549  & 5474  & 72.51$\%$ \\
          & 10000 & 2998  & 2803  & 93.50$\%$ & 10042 & 7157  & 71.27$\%$ \\
          & 20000 & 3258  & 3054  & 93.74$\%$ & 11723 & 8317  & 70.95$\%$ \\
          & 40000 & 4171  & 3858  & 92.50$\%$ & 12866 & 9200  & 71.51$\%$ \\
    \midrule
    \multirow{4}[2]{*}{\makecell{$\gamma=0.5$\\$\delta=2$}} & 4000  & 11201 & 8693  & 77.61$\%$ & 13376 & 10048 & 75.12$\%$ \\
          & 10000 & 10723 & 8906  & 83.06$\%$ & 17647 & 13311 & 75.43$\%$ \\
          & 20000 & 10652 & 9163  & 86.02$\%$ & 20226 & 15488 & 76.57$\%$ \\
          & 40000 & 12027 & 10691 & 88.89$\%$ & 21920 & 17251 & 78.70$\%$ \\
    \midrule
    \multirow{4}[2]{*}{\makecell{$\gamma=0.5$\\$\delta=4$}} & 4000  & 10202 & 9042  & 88.63$\%$ & 12254 & 10671 & 87.08$\%$ \\
          & 10000 & 13561 & 12057 & 88.91$\%$ & 16329 & 14111 & 86.42$\%$ \\
          & 20000 & 15912 & 14266 & 89.66$\%$ & 19063 & 16462 & 86.36$\%$ \\
          & 40000 & 17595 & 16085 & 91.42$\%$ & 20861 & 18302 & 87.73$\%$ \\
    \bottomrule
    \end{tabular}%
  \label{tab:AS2_steal_opt}%
  }
  \vspace{-0.5cm}
\end{table}%

\section{Notation}
\label{app:notation}
Table~\ref{tab:notation} summarizes the important notations used in this paper.

\begin{table}[h]
  \centering
  \caption{ List of notations.}
  \resizebox{0.4\textwidth}{!}{
    \begin{tabular}{cll}
    \toprule
          & Notation & Description \\
    \midrule
    \multirow{13}[2]{*}{Constant} & $T=\{ t_j \}, j \in [1,m]$ & The vocabulary of the tokens. \\
          & $S=\{S_i\}, i \in [1,n]$ & The dataset of all sentences. \\
          & $\hat{S}$, $\tilde{S}$ & The dataset of watermarked and natural sentences. \\
          & $S_i$ & Sentence $S_i$. \\
          & $s_{i,j}$ & The number of occurrences of token $t_j$ in $S_i$. \\
          & $l_i = |S_i|$ & The length of $S_i$. \\
          & $\gamma$ & The proportion of the green list in the vocabulary. \\
          & $\delta$ & The perturbation added to logits while injecting watermark. \\
          & $g_i$ & \multicolumn{1}{p{22.165em}}{Watermark threshold, the minimum green tokens required for $S_i$ when it is watermarked.} \\
          & $z^*$   & Threshold of the $z$-test score. \\
          & $K=\{k\}, p=|K|$ & The set of keys. \\
          & $e_j$, $j \in [1,m]$ & Embedding for token $t_j$. \\
          & $W=\{w_j\},  j \in [1,m]$ & The weight of each token during optimization. \\
    \midrule
    \multirow{5}[2]{*}{Variable} & $C=\{c_{j}\},  j \in [1,m]$ & Token color for each token. \\
          & \multirow{2}[0]{*}{$\hat{b}_i, \tilde{b}_i, i \in [1,n]$} & \multicolumn{1}{p{22.165em}}{Tighter bounds for watermarked and natural } \\
          &       & \multicolumn{1}{p{22.165em}}{sentences.} \\
          & $\lambda_{i} \in \{0,1\}, i \in [1,n], $ & \multicolumn{1}{p{22.165em}}{An identifier for sentence $S_i$ whether it should be considered in optimization.} \\
          & $\rho^k_i \in \{0,1\}$ & \multicolumn{1}{p{22.165em}}{An identifier for whether $k$ is suitable for sentence $S_i$.} \\
    \midrule
    \multirow{3}[2]{*}{Hyperparameter} & $\hat{\beta}, \tilde{\beta}$ & Rescale factors. \\
          & $p_u, p_l$ & The size of non-erroneous sentences. \\
          & $\mu$ & Lower bond for the size of the stolen green list in multi-key. \\
    \bottomrule
    \end{tabular}%
  \label{tab:notation}%
  \vspace{-0.5cm}
  }
\end{table}%

\end{document}